\documentclass[pre,twocolumn,aps,showpacs]{revtex4}
\newcommand{\bec}[1]{\mbox{\boldmath $ #1$}}
\usepackage{graphicx}
\begin{document}
\bigskip
\bigskip
\title{Nonlinear theory of a "shear-current" effect and mean-field
magnetic dynamos}
\author{Igor Rogachevskii}
\email{gary@menix.bgu.ac.il} \homepage{http://www.bgu.ac.il/~gary}
\author{Nathan Kleeorin}
\email{nat@menix.bgu.ac.il} \affiliation{Department of Mechanical
Engineering, The Ben-Gurion University of the Negev, \\
POB 653, Beer-Sheva 84105, Israel}
\date{\today}
\begin{abstract}
The nonlinear theory of a "shear-current" effect in a nonrotating
and nonhelical homogeneous turbulence with an imposed mean
velocity shear is developed. The ''shear-current" effect is
associated with the $\bar{\bf W} {\bf \times} \bar{\bf J}$-term in
the mean electromotive force and causes the generation of the mean
magnetic field even in a nonrotating and nonhelical homogeneous
turbulence (where $\bar{\bf W}$ is the mean vorticity and
$\bar{\bf J}$ is the mean electric current). It is found that
there is no quenching of the nonlinear "shear-current" effect
contrary to the quenching of the nonlinear $\alpha$-effect, the
nonlinear turbulent magnetic diffusion, etc. During the nonlinear
growth of the mean magnetic field, the ''shear-current" effect
only changes its sign at some value $\bar{\bf B}_\ast$ of the mean
magnetic field. The magnitude $\bar{\bf B}_\ast$ determines the
level of the saturated mean magnetic field which is less than the
equipartition field. It is shown that the background magnetic
fluctuations due to the small-scale dynamo enhance the
"shear-current" effect, and reduce the magnitude $\bar{\bf
B}_\ast$. When the level of the background magnetic fluctuations
is larger than $1/3$ of the kinetic energy of the turbulence, the
mean magnetic field can be generated due to the "shear-current"
effect for an arbitrary exponent of the energy spectrum of the
velocity fluctuations.
\end{abstract}


\maketitle

\section{Introduction}

The magnetic fields of the Sun, solar type stars, galaxies and
planets are believed to be generated by a dynamo process (see,
e.g., \cite{M78,P79,KR80,ZRS83,RSS88,S89,RS92,BB96,K99,B2000}). In
the framework of the mean-field approach, the large-scale magnetic
field $\bar{\bf B}$ is determined by the induction equation
\begin{eqnarray}
{\partial \bar{\bf B} \over \partial t} = \bec{\nabla} \times
(\bar{\bf U} {\bf \times} \bar{\bf B} + \bec{\cal E}(\bar{\bf B})
- \eta \bec{\nabla} {\bf \times} \bar{\bf B}) \;, \label{I1}
\end{eqnarray}
where $\bar{\bf U}$ is the mean velocity, $\eta$ is the magnetic
diffusion due to the electrical conductivity of fluid. The mean
electromotive force $\bec{\cal E}(\bar{\bf B}) = \langle {\bf u}
\times {\bf b} \rangle$ is given by
\begin{eqnarray}
{\cal E}_{i}(\bar{\bf B}) &=& {\alpha}_{ij}(\bar{\bf B}) \bar
B_{j} - {\eta}_{ij}(\bar{\bf B}) (\bec{\nabla} {\bf \times} \bar
{\bf B})_{j} + ({\bf V}^{\rm eff}(\bar{\bf B}) {\bf \times} \bar
{\bf B})_{i}
\nonumber\\
&& - [\bec{\delta}(\bar{\bf B}) {\bf \times} (\bec{\nabla} {\bf
\times} \bar{\bf B})]_i - {\kappa}_{ijk}(\bar{\bf B}) ({\partial
\hat B})_{jk} \;,
\label{I2}
\end{eqnarray}
where ${\bf u}$ and ${\bf b}$ are fluctuations of the velocity and
magnetic field, respectively, angular brackets denote ensemble
averaging, $({\partial \hat B})_{ij} = (\nabla_i \bar B_{j} +
\nabla_j \bar B_{i}) / 2$ is the symmetric part of the gradient
tensor of the mean magnetic field $\nabla_i \bar B_{j}$, i.e.,
$\nabla_i \bar B_{j} = (\partial \hat B)_{ij} + \varepsilon_{ijn}
(\bec{\bf \nabla} {\bf \times} \bar{\bf B})_{n} / 2 $ and
$\varepsilon_{ijk}$ is the Levi-Civita tensor. Here
$\alpha_{ij}(\bar{\bf B})$ and $ \eta_{ij}(\bar{\bf B}) $
determine the $\alpha$ effect and turbulent magnetic diffusion,
respectively, ${\bf V}^{\rm eff}(\bar{\bf B})$ is the effective
drift velocity of the magnetic field, $\kappa_{ijk}(\bar{\bf B})$
describes a contribution to the mean electromotive force related
with the symmetric parts of the gradient tensor of the mean
magnetic field, $({\partial \hat B})_{ij}$, and arises in an
anisotropic turbulence, and the $\bec{\delta}(\bar{\bf B})$-term
determines a nontrivial behavior of the mean magnetic field in an
anisotropic turbulence.

The mean magnetic field can be generated in a helical rotating
turbulence due to the $\alpha$ effect described by
${\alpha}_{ij}(\bar{\bf B}) \bar B_{j}$ term in the mean
electromotive force. When the rotation is a nonuniform, the
generation of the mean magnetic field is caused by the $\alpha
{\bf \Omega} $-dynamo. For a rotating nonhelical turbulence the
$\bec{\delta}$-term in the mean electromotive force describes the
${\bf \Omega \times J}$--effect which causes a generation of the
mean magnetic field if rotation is a nonuniform (see
\cite{R69,R72,MP82,R86,RKR03}), where ${\bf \Omega}$ is the
angular velocity and ${\bf J}$ is the mean electric current.

For a nonrotating and nonhelical turbulence the $\alpha$ effect
and the ${\bf \Omega \times J}$--effect vanish. However, a mean
magnetic field can be generated in a nonrotating and nonhelical
turbulence with an imposed mean velocity shear due to the
''shear-current" effect \cite{RK03}, described by the
$\bec{\delta}$-term in the mean electromotive force. In order to
elucidate the physics of the ''shear-current" effect, we compare
the $\alpha$-effect in the $\alpha {\bf \Omega} $-dynamo with the
$\bec{\delta}$-term caused by the ''shear-current" effect. The
$\alpha$-term  in the mean electromotive force which is
responsible for the generation of the mean magnetic field, reads $
{\cal E}^\alpha_i \equiv \alpha \bar B_i \propto - ({\bf \Omega}
\cdot {\bf \Lambda}) \bar B_i $ (see, e.g., \cite{KR80,RKR03}),
where $ {\bf \Lambda} = \bec{\nabla} \langle {\bf u}^2 \rangle /
\langle {\bf u}^2 \rangle $ determines one of the inhomogeneities
of the turbulence. The $\bec{\delta}$-term in the electromotive
force caused by the ''shear-current" effect is given by $ {\cal
E}^\delta_i \equiv - (\bec{\delta} {\bf \times} (\bec{\nabla} {\bf
\times} \bar{\bf B}))_i \propto - (\bar{\bf W} \cdot \bec{\nabla})
\bar B_i $  (see Eq.~(\ref{K10}) below, and \cite{RK03}), where
the $\bec{\delta}$-term is proportional to the mean vorticity
$\bar{\bf W} = \bec{\nabla} {\bf \times} \bar {\bf U}$. The mean
vorticity $\bar{\bf W}$ in the ''shear-current" dynamo plays a
role of a differential rotation and an inhomogeneity of the mean
magnetic field plays a role of the inhomogeneity of turbulence.
During the generation of the mean magnetic field in both cases (in
the $\alpha {\bf \Omega} $-dynamo and in the ''shear-current"
dynamo), the mean electric current along the original mean
magnetic field arises. The $\alpha$-effect is related with the
hydrodynamic helicity $ \propto ({\bf \Omega} \cdot {\bf \Lambda})
$ in an inhomogeneous turbulence. The deformations of the magnetic
field lines are caused by upward and downward rotating turbulent
eddies in the $\alpha {\bf \Omega} $-dynamo. Since the turbulence
is inhomogeneous (which breaks a symmetry between the upward and
downward eddies), their total effect on the mean magnetic field
does not vanish and it creates the mean electric current along the
original mean magnetic field.

In a turbulent flow with an imposed mean velocity shear, the
inhomogeneity of the original mean magnetic field breaks a
symmetry between the influence of upward and downward turbulent
eddies on the mean magnetic field. The deformations of the
magnetic field lines in the ''shear-current" dynamo are caused by
upward and downward turbulent eddies which result in the mean
electric current along the mean magnetic field and produce the
magnetic dynamo.

The ''shear-current" effect was studied in \cite{RK03} in a
kinematic approximation. Kinematic dynamo models predict a field
that grows without limit, and they give no estimate of the
magnitude for the generated magnetic field. In order to find the
magnitude of the field, the nonlinear effects which limit the
field growth must be taken into account.

The main goal of this study is to develop a nonlinear theory of
the ''shear-current" effect. We demonstrated that the nonlinear
''shear-current" effect is very important nonlinearity in a
mean-field dynamo. During the nonlinear growth of the mean
magnetic field, the ''shear-current" effect changes its sign, but
there is no quenching of this effect contrary to the quenching of
the nonlinear $\alpha$-effect, the nonlinear turbulent magnetic
diffusion, etc. The nonlinear ''shear-current" effect determines
the level of the saturated mean magnetic field. This paper is
organized as follows. First, we discuss qualitatively a mechanism
for the "shear-current" effect (Section II). In Section III we
formulate the governing equations, the assumptions and the
procedure of the derivation of the nonlinear mean electromotive
force in a turbulent flow with a mean velocity shear. In Section
IV we analyze the coefficients defining the mean electromotive
force for a shear-free turbulence and for a sheared turbulence,
and consider the implications of the obtained results to the
mean-field magnetic dynamo. The nonlinear saturation of the mean
magnetic field and astrophysical applications of the obtained
results are discussed in Section V.

\section{The ''shear-current" effect}

In order to describe the ''shear-current" effect we need to
determine the mean electromotive force. The general form of the
mean electromotive force in a turbulent flow with a mean velocity
shear can be obtained even from simple symmetry reasoning. Indeed,
the mean electromotive force can be written in the form:
\begin{eqnarray}
{\cal E}_{i} = a_{ij} \, \bar B_{j} + b_{ijk} \, \bar B_{j,k} \;,
\label{A23}
\end{eqnarray}
where $\bar B_{j,i} = \nabla_i \bar B_{j}$ and we neglected terms
$\sim O(\nabla^2 \bar B_{k})$. Following to \cite{R80} we rewrite
Eq.~(\ref{A23}) for the mean electromotive force in the form of
Eq.~(\ref{I2}) with
\begin{eqnarray}
\alpha_{ij}(\bar{\bf B}) &=& {1 \over 2}(a_{ij} + a_{ji}) \;,
\quad V^{\rm eff}_k(\bar{\bf B}) = {1 \over 2} \varepsilon_{kji}
\, a_{ij} \;,
\label{A25} \\
\eta_{ij}(\bar{\bf B}) &=& {1 \over 4}(\varepsilon_{ikp} \,
b_{jkp} + \varepsilon_{jkp} \, b_{ikp}) \;, \quad \delta_{i} = {1
\over 4} (b_{jji} - b_{jij}) \;,
\nonumber \\
\label{A26}\\
\kappa_{ijk}(\bar{\bf B}) &=& - {1 \over 2}(b_{ijk} + b_{ikj}) \;,
\label{A27}
\end{eqnarray}
where we used an identity $ \bar B_{j,i} = (\partial \hat B)_{ij}
+ \varepsilon_{ijn} (\bec{\bf \nabla} {\bf \times} \bar{\bf
B})_{n} / 2 $. Note that the separation of terms in
Eqs.~(\ref{A23})-(\ref{A27}) is not unique, because a gradient
term can always be added to the electromotive force. Let us
consider a homogeneous, nonrotating and nonhelical turbulence.
Then in the kinematic approximation the tensor $a_{ij}$ vanishes.
This implies that $\alpha_{ij} = 0 $ and $V^{\rm eff}_{k} = 0 .$
The mean electromotive force $\bec{\cal E}$ is a true vector,
whereas the mean magnetic field $\bar{\bf B}$ is a pseudo-vector.
Thus, the tensor $b_{ijk}$ is a pseudo-tensor (see
Eq.~(\ref{A23})). For homogeneous, isotropic and nonhelical
turbulence the tensor $b_{ijk} = \eta_{_{T}} \varepsilon_{ijk} ,$
where $\eta_{_{T}}= u_0 l_0 / 3$ is the coefficient of isotropic
turbulent magnetic diffusion, $u_0$ is the characteristic
turbulent velocity in the maximum scale of turbulent motions
$l_0$. In a turbulent flow with an imposed mean velocity shear,
the tensor $b_{ijk}$ depends on the true tensor $\nabla_j \bar
U_{i} .$ In this case turbulence is anisotropic. The tensor $
\nabla_j \bar U_{i} $ can be written as a sum of the symmetric and
antisymmetric parts, i.e., $ \nabla_j \bar U_{i} = (\partial \hat
U)_{ij} - (1/2) \varepsilon_{ijk} \, \bar W_{k} ,$ where $
(\partial \hat U)_{ij} = (\nabla_i \bar U_{j} + \nabla_j \bar
U_{i}) / 2 $ is the true tensor and the mean vorticity $\bar{\bf
W}$ is a pseudo-vector. Now we take into account the effect which
is linear in $ \nabla_j \bar U_{i} .$ Thus, the pseudo-tensor
$b_{ijk}$  in the kinematic approximation has the following form
\begin{eqnarray}
b_{ijk} &=& \eta_{_{T}} \varepsilon_{ijk} + l_0^2 \, [D_1 \,
\varepsilon_{ijm} (\partial \hat U)_{mk} + D_2 \,
\varepsilon_{ikm} (\partial \hat U)_{mj}
\nonumber\\
&& + D_3 \, \varepsilon_{jkm} (\partial \hat U)_{mi} + D_4 \,
\delta_{ij} \bar W_{k} + D_5 \, \delta_{ik} \bar W_{j}] \;,
\label{K2}
\end{eqnarray}
where $ D_k $ are the unknown coefficients, $\delta_{ij}$ is the
Kronecker tensor, and the term $\propto \delta_{jk} \bar W_{i}$
vanishes since $ \bec{\nabla} \cdot \bar{\bf B} = 0 $ (see
Eq.~(\ref{A23})). Using Eqs.~(\ref{A25})-(\ref{A27}) we determine
the turbulent coefficients defining the mean electromotive force
for a homogeneous and nonhelical turbulence with a mean velocity
shear:
\begin{eqnarray}
\eta_{ij} &=& \eta_{_{T}} \, \delta_{ij} - 2 \, l_0^2 \, \eta_{0}
\, (\partial \hat U)_{ij} \,, \quad \bec{\delta} = l_0^2 \,
\delta_0 \, \bar{\bf W} \,,
\label{K10}\\
\kappa_{ijk} &=& l_0^2 \, [\kappa_{1} \, \delta_{ij} \, \bar W_k +
\kappa_{2} \, \varepsilon_{ijm} \, (\partial \hat U)_{mk}] \,,
\label{K11}
\end{eqnarray}
where $\eta_{0} = (D_1 - D_2 - 2 D_3) / 4 \,,$ $\, \delta_0 = (D_4
- D_5) / 2 \,,$ $ \kappa_{1} = - (D_4 + D_5) \,$ and $\,
\kappa_{2} = - (D_1 + D_2) $. The second term in the tensor
$\eta_{ij}$ describes an anisotropic part of turbulent magnetic
diffusion caused by the mean velocity shear, while the first term
in the tensor $\eta_{ij}$ is the isotropic contribution to
turbulent magnetic diffusion. The $\bec{\delta}$ term for the mean
electromotive force describes the ''shear-current" effect which
can cause the mean-field magnetic dynamo. Indeed, consider a
homogeneous divergence-free turbulence with a mean velocity shear,
$ \bar{\bf U} = (0, Sx, 0)$ and $ \bar{\bf W} = (0,0,S) .$ Let us
study a simple case when the mean magnetic field is $ \bar{\bf B}
= (\bar B_x(z), \bar B_y(z), 0) .$ The mean magnetic field in the
kinematic approximation is determined by
\begin{eqnarray}
{\partial \bar B_x \over \partial t} &=& - S \, l_0^2 \, \sigma_0
\, \bar B''_y + \eta_{_{T}} \, \bar B''_x  \;,
\label{E2}\\
{\partial \bar B_y \over \partial t} &=& S \, \bar B_x +
\eta_{_{T}} \, \bar B''_y  \;,
\label{E3}
\end{eqnarray}
where $\bar B'' = \partial^2 \bar B / \partial z^2 $ and
\begin{eqnarray}
\sigma_0 = \delta_0 - \eta_0 - {\kappa_1 \over 2} - {\kappa_2
\over 4} = {1 \over 2}(D_2 + D_3 + 2 D_4) \; . \label{E30}
\end{eqnarray}
In Eq.~(\ref{E3}) we took into account that the characteristic
spatial scale $ L_B $ of the mean magnetic field variations is
much larger than the maximum scale of turbulent motions $ l_0 .$
Equation~(\ref{E30}) was obtained in \cite{RK03} in the kinematic
approximation. A solution of Eqs.~(\ref{E2}) and~(\ref{E3}) we
seek for in the form $ \propto \exp(\gamma \, t + i K_z \, z) ,$
where the growth rate $\gamma$ of the magnetic dynamo instability
is given by
\begin{eqnarray}
\gamma = S \, l_0 \, K_z \, \sqrt{\sigma_0} - \eta_{_{T}} \, K_z^2
\; . \label{K6}
\end{eqnarray}
The first term $ (\propto S \bar B_x) $ in RHS of Eq.~(\ref{E3})
describes the shear motions. This effect is similar to the
differential rotation because $ \bec{\nabla} {\bf \times}
(\bar{\bf U} {\bf \times} \bar{\bf B}) = S \bar B_x {\bf e}_y .$
The magnetic dynamo instability is determined by a coupling
between the components of the mean magnetic field. In particular,
the inhomogeneous magnetic field $ \bar B_y $ generates the field
$ \bar B_x $ due to the ''shear-current" effect (described by the
first term in RHS of Eq.~(\ref{E2})). This is similar to the
$\alpha$ effect. On the other hand, the field $ \bar B_x $
generates the field $ \bar B_y $ due to the pure shear effect
(described by the first term in RHS of Eq.~(\ref{E3})), like the
differential rotation in $\alpha \Omega$-dynamo. It follows from
Eqs.~(\ref{E2}) and~(\ref{E3}) that for the ''shear-current"
dynamo, $\bar B_x / \bar B_y \sim l_0 / L_B \ll 1 .$ Note that in
the $\alpha \Omega$-dynamo, the poloidal component of the mean
magnetic field is much smaller than the toroidal field.

The magnetic dynamo instability due to the ''shear-current" effect
is different from that for $\alpha \Omega$-dynamo. Indeed, the
dynamo mechanism due to the ''shear-current" effect acts even in
homogeneous nonhelical turbulence, while the alpha effect vanishes
for homogeneous turbulence.

The ''shear-current" effect was studied in \cite{RK03} in the
kinematic approximation using two different methods: the
$\tau$--approximation (the Orszag third-order closure procedure)
and the stochastic calculus (the path integral representation of
the solution of the induction equation, Feynman-Kac formula and
Cameron-Martin-Girsanov theorem). The $\bec{\delta}$--term in the
electromotive force which is responsible for the ''shear-current"
effect was also independently found in \cite{RS02} in a problem of
a screw dynamo using the modified second-order correlation
approximation.

\section{The governing equations and the procedure of the
derivation of the nonlinear effects}

Now let us develop a nonlinear theory of the ''shear-current"
effect. In order to derive equations for the nonlinear
coefficients defining the mean electromotive force in a
homogeneous turbulence with a mean velocity shear, we will use a
mean field approach in which the magnetic and velocity fields are
divided into the mean and fluctuating parts, where the fluctuating
parts have zero mean values. The procedure of the derivation of
equation for the nonlinear mean electromotive force is as follows
(for details, see Appendix A). We consider the case of large
hydrodynamic and magnetic Reynolds numbers. The momentum equation
and the induction equation for the turbulent fields are given by
\begin{eqnarray}
{\partial {\bf u}(t,{\bf x}) \over \partial t} &=& - {\bec{\nabla}
p_{\rm tot} \over \rho} + {1 \over \mu \rho} [({\bf b} \cdot
\bec{\nabla}) \bar{\bf B} + (\bar{\bf B} \cdot \bec{\nabla}){\bf
b}]
\nonumber \\
&& - (\bar{\bf U} \cdot \bec{\nabla}) {\bf u} - ({\bf u} \cdot
\bec{\nabla}) \bar{\bf U} + {\bf u}^N + {\bf F} \,,
\label{B1} \\
{\partial {\bf b}(t,{\bf x}) \over \partial t} &=& (\bar{\bf B}
\cdot \bec{\nabla}){\bf u} - ({\bf u} \cdot \bec{\nabla}) \bar{\bf
B} - (\bar{\bf U} \cdot \bec{\nabla}) {\bf b}
\nonumber \\
&& + ({\bf b} \cdot \bec{\nabla}) \bar{\bf U} + {\bf b}^N \,,
\label{B2}
\end{eqnarray}
where ${\bf u}$ and ${\bf b}$ are fluctuations of velocity and
magnetic field, respectively, $\bar{\bf B}$ is the mean magnetic
field, $\bar{\bf U}$ is the mean velocity field, $ \rho $ is the
fluid density, $\mu$ is the magnetic permeability of the fluid, $
{\bf F} $ is a random external stirring force, ${\bf u}^{N}$ and
${\bf b}^{N}$ are the nonlinear terms which include the molecular
dissipative terms, $ p_{\rm tot} = p + \mu^{-1} \,(\bar{\bf B}
\cdot {\bf b}) $ are the fluctuations of total pressure, $p$ are
the fluctuations of fluid pressure. The velocity ${\bf u}$
satisfies to the equation: $\bec{\nabla} \cdot {\bf u} = 0 .$
Hereafter we omit $\mu$ in equations, i.e., we include
$\mu^{-1/2}$ in the definition of magnetic field. We study the
effect of a mean velocity shear on the mean electromotive force.

Using Eqs.~(\ref{B1})-(\ref{B2}) written in a Fourier space we
derive equations for the correlation functions of the velocity
field $f_{ij}=\langle u_i u_j \rangle $, the magnetic field
$h_{ij}=\langle b_i b_j \rangle $ and the cross-helicity
$g_{ij}=\langle b_i u_j \rangle $. The equations for these
correlation functions are given by Eqs. (\ref{B6})-(\ref{B8}) in
Appendix A. We split the tensors $ f_{ij}, h_{ij}$ and $g_{ij}$
into nonhelical, $f_{ij},$ and helical, $f_{ij}^{(H)},$ parts. The
helical part of the tensor $h_{ij}^{(H)}$ for magnetic
fluctuations depends on the magnetic helicity, and it is
determined by the dynamic equation which follows from the magnetic
helicity conservation arguments (see, {\rm e.g.,}
\cite{KR82,KRR94,GD94,KR99,KMRS2000,BB02}). The characteristic
time of evolution of the nonhelical part of the magnetic tensor
$h_{ij}$ is of the order of the turbulent correlation time $
\tau_{0} = l_{0} / u_{0} ,$ while the relaxation time of the
helical part of the magnetic tensor $ h_{ij}^{(H)} $ is of the
order of $ \tau_{0} \,{\rm Rm} $ (see, {\em e.g.,} \cite{KR99}),
where ${\rm Rm} = l_0 u_{0} / \eta \gg 1 $ is the magnetic
Reynolds number, $u_{0}$ is the characteristic turbulent velocity
in the maximum scale $l_{0}$ of turbulent motions.

Then we split the nonhelical parts of the correlation functions $
f_{ij}, h_{ij}$ and $g_{ij}$ into symmetric and antisymmetric
tensors with respect to the wave vector ${\bf k}$, {\em e.g.,} $
f_{ij} = f_{ij}^{(s)} + f_{ij}^{(a)} ,$ where the tensors $
f_{ij}^{(s)} = [f_{ij}({\bf k}) + f_{ij}(-{\bf k})] / 2 $
describes the symmetric part of the tensor and $ f_{ij}^{(a)} =
[f_{ij}({\bf k}) - f_{ij}(-{\bf k})] / 2 $ determines the
antisymmetric part of the tensor.

Equations for the second moments contain higher moments and a
problem of closing the equations for the higher moments arises.
Various approximate methods have been proposed for the solution of
problems of this type (see, {\em e.g.,} \cite{MY75,Mc90,O70}). The
simplest procedure is the $ \tau $-approximation, which is widely
used in the theory of kinetic equations, in passive scalar
turbulence and magnetohydrodynamic turbulence (see, {\em e.g.,}
\cite{O70,PFL76,KRR90,KMR96,RK2000,KR03,BK04}). This procedure
allows us to express the deviations of the third moments from the
background turbulence in terms of the corresponding deviations of
the second moments, e.g.,
\begin{eqnarray}
f_{ij}^N - f_{ij}^{N(0)} &=& - (f_{ij} - f_{ij}^{(0)}) / \tau (k)
\;,
\label{A1}
\end{eqnarray}
where the term $ f_{ij}^N $ is related with the third moment (see
Appendix A). The superscript $ {(0)} $ corresponds to the
background turbulence (with $ \bar {\bf B} = 0)$, and $ \tau (k) $
is the characteristic relaxation time of the statistical moments.
We applied the $ \tau $-approximation only for the nonhelical part
$h_{ij}$ of the tensor of magnetic fluctuations.

In this study we consider an intermediate nonlinearity which
implies that the mean magnetic field is not enough strong in order
to  affect the correlation time of turbulent velocity field. The
theory for a very strong mean magnetic field can be corrected
after taking into account a dependence of the correlation time of
the turbulent velocity field on the mean magnetic field. We assume
that the characteristic time of variation of the mean magnetic
field $\bar{\bf B}$ is substantially larger than the correlation
time $\tau(k)$ for all turbulence scales (which corresponds to the
mean-field approach). This allows us to get a stationary solution
for the equations for the second moments $ f_{ij}, \, h_{ij} $ and
$g_{ij} $. For the integration in $ {\bf k} $-space of these
second moments we have to specify a model for the background
turbulence (with $ \bar{\bf B} = 0)$. We use the following model
for the background homogeneous and isotropic turbulence:
$f_{ij}^{(0)}({\bf k}) = \langle {\bf u}^2 \rangle^{(0)} \, W(k)
\, (\delta_{ij} - k_{ij}) ,$ $\, h_{ij}^{(0)}({\bf k}) = \langle
{\bf b}^2 \rangle^{(0)} \, W(k) \, (\delta_{ij} - k_{ij}) $ and
$\, g_{ij}^{(0)}({\bf k}) = 0 ,$ where $ \, k_{ij} = k_{i} k_{j} /
k^{2} ,$ $\, W(k) = - (d \bar \tau(k) / dk) / 8 \pi k^{2} ,$ $ \,
\tau(k) = 2 \tau_{0} \bar \tau(k) ,$ $ \, \bar \tau(k) = (k /
k_{0})^{1-q} ,$ $\, 1 < q < 3 $ is the exponent of the kinetic
energy spectrum (e.g., $ q = 5/3 $ for Kolmogorov spectrum), $
k_{0} = 1 / l_{0} ,$ and $\tau_{0} = l_{0} / u_{0} $, $\, \int
f_{ij}^{(0)}({\bf k}) \,d {\bf k} = (\langle {\bf u}^2
\rangle^{(0)} / 3) \delta_{ij} $ and $\int h_{ij}^{(0)}({\bf k})
\,d {\bf k} = (\langle {\bf b}^2 \rangle^{(0)} / 3) \delta_{ij} .$

Using the derived equations for the second moments $ f_{ij}, \,
h_{ij} $ and $g_{ij} $ we calculate the mean electromotive force $
{\cal E}_{i} = \int \tilde {\cal E}_{i}({\bf k}) \,d {\bf k} ,$
where $ \tilde{\cal E}_{i}({\bf k}) = \varepsilon_{imn}
g_{nm}^{(s)}({\bf k}) .$ For a turbulence with a mean velocity
shear the coefficients defining the mean electromotive force are
the sum of contributions arising from a shear-free turbulence and
sheared turbulence (see Section IV).

\section{The nonlinear mean-field dynamo in a turbulence
with a mean velocity shear}

First, let us consider a shear-free nonrotating homogeneous and
nonhelical turbulence. Using Eqs.~(\ref{A23})-(\ref{A27})
and~(\ref{L1})-(\ref{L2}) we derive equations for the mean
electromotive force. The coefficients defining the mean
electromotive force for a shear-free turbulence in a dimensionless
form are given by
\begin{eqnarray}
\alpha_{ij}^{(m)} &=& \alpha^{(m)}(\bar{\bf B}) \, \delta_{ij} \;,
\label{L20}\\
{\bf V}^{\rm eff} &=& {\bf V}_{A}(\bar{B}) +  \tilde \eta(\bar{B})
 \, {(\bar{\bf B} \cdot \bec{\nabla}) \bar{\bf
B} \over \bar{\bf B}^2} \;,
\label{B40}\\
\eta_{ij} &=& \eta_{_{A}}(\bar{B}) \, \delta_{ij} \;,
\label{B41}
\end{eqnarray}
where $\tilde \eta(\bar{B}) = - (1/2) (1 + \epsilon) A_{2}^{(1)}(4
\bar{B}) $ , the functions $\eta_{_{A}}(\bar{B})$ and ${\bf
V}_{A}(\bar{B})$ are determined by Eqs.~(\ref{B60}) and
~(\ref{B62}) respectively, the functions $A_{k}^{(1)}(y)$ are
determined by Eqs.~(\ref{P20}) in Appendix C, the parameter
$\epsilon = \langle {\bf b}^2 \rangle^{(0)} / \langle {\bf u}^2
\rangle^{(0)}$ is the ratio of the magnetic and kinetic energies
in the background turbulence. The function $\alpha^{(m)}(\bar{\bf
B}) = \chi^{(c)}(\bar{\bf B}) \, \phi^{(m)}(\sqrt{8} \bar B)$ is
the magnetic part of the $\alpha$-effect, where $\phi^{(m)}(y) =
(3 / y^{2}) (1 - \arctan y / y) $ is the quenching function of the
magnetic part of the $\alpha$-effect (see \cite{FB99,RK2000}), and
the dimensionless function $\chi^{(c)}(\bar {\bf B}) = (\tau / 3
\mu \rho \, u_0) \langle {\bf b} \cdot (\bec{\nabla} {\bf \times}
{\bf b}) \rangle$. The function $\chi^{(c)}(\bar {\bf B})$ is
determined by the dynamic equation which follows from the magnetic
helicity conservation arguments (see, {\rm e.g.,}
\cite{KR82,KRR94,GD94,KR99,KMRS2000,BB02}). Note that in a
homogeneous and nonhelical background turbulence the hydrodynamic
part, $\alpha_{ij}^{(u)}$,  of the $\alpha$ effect vanishes. In a
turbulence without a uniform rotation or a mean velocity shear,
the $\bec{\delta}(\bar{\bf B})$-term and the
${\kappa}_{ijk}(\bar{\bf B})$-term in the mean electromotive force
vanish.

We adopt here the dimensionless form of the mean dynamo equations;
in particular, length is measured in units of $L$, time is
measured in units of $ L^{2} / \eta_{_{T}} $ and $\bar{\bf B}$ is
measured in units of the equipartition energy $\bar B_{\rm eq} =
\sqrt{\mu \rho} \, u_0 $, the nonlinear turbulent magnetic
diffusion coefficients are measured in units of the characteristic
value of the turbulent magnetic diffusivity $ \eta_{_{T}} = l_0
u_{0} / 3 $. Note that $L \sim L_B$, where $L_B$ is the
characteristic scale of the mean magnetic field variations.

Now we consider a small-scale homogeneous turbulence with a mean
velocity shear, $ \bar{\bf U} = S \, x \, {\bf e}_y $ and $
\bar{\bf W} = S \, {\bf e}_z .$ In cartesian coordinates the mean
magnetic field, $\bar{\bf B} = B(x,z) \, {\bf e}_y + \bec{\nabla}
{\bf \times} [A(x,z) \, {\bf e}_y]$, is determined by the
dimensionless dynamo equations
\begin{eqnarray}
{\partial A \over \partial t} &=& \biggl({l_0 \over L} \biggr)^2
\, S_\ast \, \sigma_0(\bar{B}) \, (\hat {\bf W} \cdot
\bec{\nabla}) \, B + \alpha^{(m)}(\bar{\bf B}) \, B
\nonumber\\
&& - ({\bf V}_{A}(\bar{B}) \cdot \bec{\nabla}) \, A +
\eta_{_{A}}(\bar{B}) \, \Delta A \;,
\label{F11} \\
{\partial B \over \partial t} &=& - S_\ast \, (\hat {\bf W} \cdot
\bec{\nabla}) \, A + \bec{\nabla} \cdot (\eta_{_{B}}(\bar{B})
\bec{\nabla} B) \;, \label{F12}
\end{eqnarray}
where $S_\ast = S \, L^2 / \eta_{_{T}}$,  and $\hat {\bf W} =
\bar{\bf W} / \bar{W}$, the function $\sigma_0(\bar{B})$ is
determined below, the nonlinear turbulent magnetic diffusion
coefficients and the nonlinear drift velocities of the mean
magnetic field are given by
\begin{eqnarray}
\eta_{_{A}}(\bar{B}) &=& A_{1}^{(1)}(4 \bar B) + A_{2}^{(1)}(4
\bar B) \;,
\label{B60} \\
\eta_{_{B}}(\bar{B}) &=& A_{1}^{(1)}(4 \bar B) + 3 (1 - \epsilon)
\biggl[A_{2}^{(1)}(4 \bar B)
\nonumber\\
&&- {1 \over 2\pi} \bar A_{2}(16 \bar B^2)\biggr] \;,
\label{B61} \\
{\bf V}_{A}(\bar{B}) &=& - {{\bf \Lambda}^{(B)} \over 2} \biggl[(2
- 3 \epsilon) A_{2}^{(1)}(4 \bar B)
\nonumber\\
&& - (1 - \epsilon) {3 \over 2 \pi} \bar A_{2}(16 \bar B^2)\biggr]
\;,
\label{B62}
\end{eqnarray}
where ${\bf \Lambda}^{(B)} = \bec{\nabla} \bar{B}^2 / \bar{B}^2 $,
the parameter $0 \leq \epsilon \leq 1 $, the functions $\bar
A_{k}(y)$ and $A_{k}^{(1)}(y)$ are determined by Eqs.~(\ref{P22})
and ~(\ref{P20}) in Appendixes B and C. For derivations
Eqs.~(\ref{B60})-(\ref{B62}) we used Eqs.~(\ref{B40}) and
(\ref{B41}). Note that in Eqs.~(\ref{B60})-(\ref{B62}) we
neglected small contributions $\sim O[(l_0 / L)^2]$ caused by the
mean velocity shear. The nonlinear turbulent magnetic diffusion
coefficients $\eta_{_{A}}$ and $\eta_{_{B}}$ and the nonlinear
effective drift velocity $V_A$ of mean magnetic field for
different value of the parameter $\epsilon$ are shown in FIGS.
1-2. The background magnetic fluctuations caused by the
small-scale dynamo result in increase of the nonlinear turbulent
magnetic diffusion coefficient $\eta_{_{B}}$, and they do not
affect the nonlinear turbulent magnetic diffusion coefficient
$\eta_{_{A}}$ (see FIG. 1). On the other hand, the background
magnetic fluctuations strongly affect the nonlinear effective
drift velocity $V_A$ of mean magnetic field. In particular, when
$\epsilon > 1/2$, the velocity $V_A$ is negative (i.e., it is
diamagnetic velocity) which causes a drift of the magnetic field
components $\bar B_x$ and $\bar B_z$  from the regions with a high
intensity of the mean magnetic field $\bar B$. When $0 < \epsilon
< 1/2$, the effective drift velocity $V_A$ is paramagnetic
velocity for a weak mean magnetic field (see FIG. 2). For strong
field, $\bar B > \bar B_{\rm eq} / 2$, the effective drift
velocity $V_A$ is diamagnetic for an arbitrary level of the
background magnetic fluctuations.

\begin{figure}
\centering
\includegraphics[width=8cm]{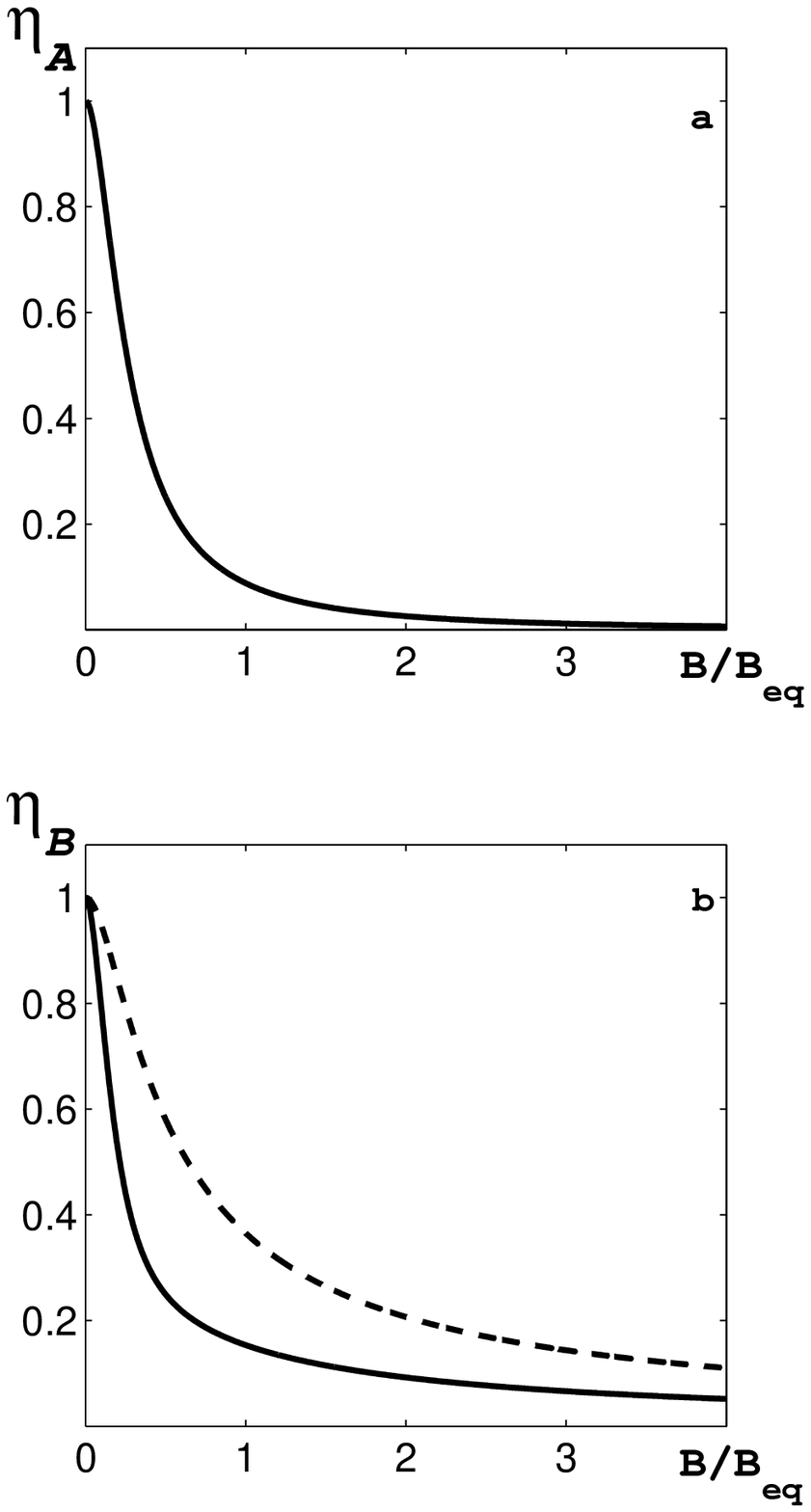}
\caption{\label{Fig1} The nonlinear turbulent magnetic diffusion
coefficients $\eta_{_{A}}$ (FIG. 1a) and $\eta_{_{B}}$ (FIG. 1b)
for $\epsilon=0$ (solid) and $\epsilon=1$ (dashed). The function
$\eta_{_{A}}$ is independent of the parameter $\epsilon$. The
nonlinear turbulent magnetic diffusion coefficients are measured
in units of the characteristic value of the turbulent magnetic
diffusivity $\eta_{_{T}} = l_0 u_{0} / 3$.}
\end{figure}

\begin{figure}
\centering
\includegraphics[width=8cm]{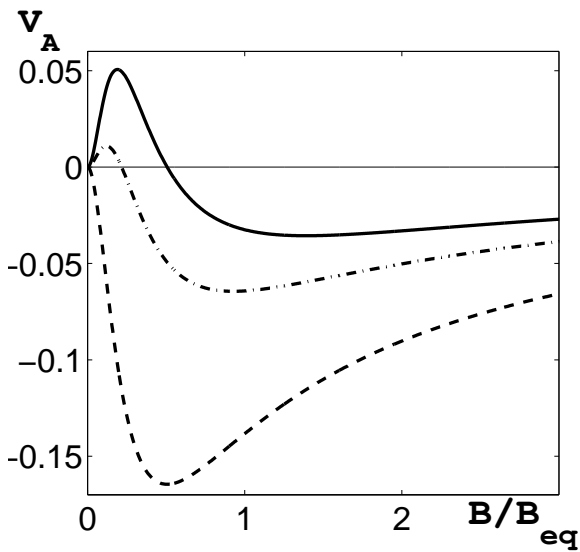}
\caption{\label{Fig2} The nonlinear effective drift velocity $V_A$
of mean magnetic field for $\epsilon=0$ (solid); $\epsilon=0.3$
(dashed-dotted); $\epsilon=1$ (dashed). The velocity $V_A$ is
measured in units of $\eta_{_{T}} / L$.}
\end{figure}

The asymptotic formulas for the magnetic part of the
$\alpha$-effect, the nonlinear turbulent magnetic diffusion
coefficients, and the nonlinear drift velocity of the mean
magnetic field for $\bar{B} \ll \bar{B}_{\rm eq} / 4$ are given by
\begin{eqnarray*}
\alpha_{ij}^{(m)}(\bar{\bf B}) &=& \chi^{(c)}(\bar{\bf B})
\biggr(1 - {3 \beta^{2} \over 5}\biggr) \delta_{ij} \;,
\\
\eta_{_{A}}(\bar{B}) &=& 1 - {12 \over 5} \, \beta^{2} \;, \quad
\eta_{_{B}}(\bar{B}) = 1 - {4 \over 5} \, (5 - 4\epsilon) \,
\beta^{2} \;,
\\
{\bf V}_{A}(\bar{B}) &=& {4 \over 5} \, (1 - 2 \epsilon) \,
\beta^{2} \, {\bf \Lambda^{(B)}} \;,
\end{eqnarray*}
and for $\bar{B} \gg \bar{B}_{\rm eq} / 4$ they are given by
\begin{eqnarray*}
\alpha_{ij}^{(m)}(\bar{\bf B}) &=& \chi^{(c)}(\bar{\bf B}) \, {3
\pi \over 2 \beta^{2}} \, \delta_{ij} \;,
\\
\eta_{_{A}}(\bar{B}) &=&  {1 \over \beta^2} \;, \quad
\eta_{_{B}}(\bar{B}) =  {2 (1 + \epsilon) \over 3 \beta} \;,
\\
{\bf V}_{A}(\bar{B}) &=& -  {1 + \epsilon \over 3 \beta} {\bf
\Lambda^{(B)}} \;,
\end{eqnarray*}
where $ \beta = \sqrt{8} \bar B $.

\begin{figure}
\centering
\includegraphics[width=8cm]{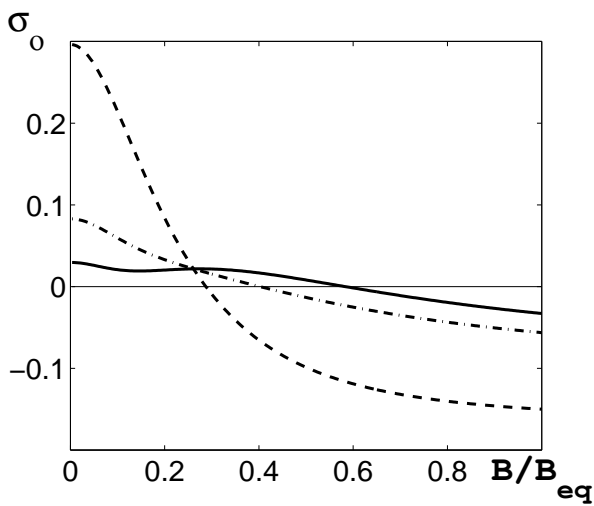}
\caption{\label{Fig3} The dimensionless nonlinear coefficient
$\sigma_0(\bar{B})$ defining the ''shear-current'' effect for
different values of the parameter $\epsilon$: $\, \, \,
\epsilon=0$ (solid); $\epsilon = 0.2 $ (dashed-dotted);
$\epsilon=1$ (dashed).}
\end{figure}

The nonlinear coefficient $\sigma_0(\bar{B})$ defining the
''shear-current'' effect is determined by Eqs.~(\ref{F1}) in
Appendix A. The nonlinear dependence of the parameter
$\sigma_0(\bar{B})$ is shown in FIG.~3 for different values of the
parameter $\epsilon$. The background magnetic fluctuations caused
by the small-scale dynamo and described by the parameter
$\epsilon$, increase the parameter $\sigma_0(\bar{B})$. Note that
the parameter $\sigma_0(\bar{B})$ is determined by the
contributions from the $\bec{\delta}(\bar{\bf B})$-term, the
${\eta}_{ij}(\bar{\bf B})$-term and the ${\kappa}_{ijk}(\bar{\bf
B})$-term in the mean electromotive force. The asymptotic formula
for the parameter $\sigma_0(\bar{B})$ for a weak mean magnetic
field $\bar{B} \ll \bar{B}_{\rm eq} / 4$ is given by
\begin{eqnarray}
\sigma_0(\bar{B}) = {4 \over 45} \, (2 - q + 3 \epsilon) \;,
\label{F2}
\end{eqnarray}
where $q$ is the exponent of the energy spectrum of the background
turbulence. In Eq.~(\ref{F2}) we neglected small contribution
$\sim O[(4 \bar{B} / \bar{B}_{\rm eq})^2]$. Equation~(\ref{F2}) is
in agreement with that obtained in \cite{RK03} where the case of a
weak mean magnetic field and $\epsilon=0$ was considered. Thus,
the mean magnetic field is generated due to the ''shear-current"
effect, when the exponent of the energy spectrum of the velocity
fluctuations is
\begin{eqnarray*}
q < 2 + 3 \epsilon \; .
\end{eqnarray*}
Note that the parameter $q$ varies in the range $1 < q < 3$.
Therefore, when the level of the background magnetic fluctuations
caused by the small-scale dynamo is larger than $1/3$ of the
kinetic energy of the velocity fluctuations, the mean magnetic
field can be generated due to the "shear-current" effect for an
arbitrary exponent $q$ of the energy spectrum of the velocity
fluctuations. For the Kolmogorov turbulence, i.e., when the
exponent of the energy spectrum of the background turbulence
$q=5/3$, the parameter $\sigma_0(\bar{B})$ for $\bar{B} \ll
\bar{B}_{\rm eq} /4$ is given by
\begin{eqnarray}
\sigma_0(\bar{B}) = {4 \over 135} \, (1 + 9 \epsilon) \;,
\label{F3}
\end{eqnarray}
and for $\bar{B} \gg \bar{B}_{\rm eq} / 4$ the parameter
$\sigma_0(\bar{B})$ is
\begin{eqnarray}
\sigma_0(\bar{B}) = - {11 \over 135} \, (1 + \epsilon) \; .
\label{F4}
\end{eqnarray}
In Eq.~(\ref{F4}) we neglected small contribution $\sim
O(\bar{B}_{\rm eq} / 4 \bar{B})$. It is seen from
Eqs.~(\ref{F2})-(\ref{F4}) that the nonlinear coefficient
$\sigma_0(\bar{B})$ defining the ''shear-current'' effect changes
its sign at some value of the mean magnetic field
$\bar{B}=\bar{B}_\ast$. For instance, $\bar{B}_\ast = 0.6
\bar{B}_{\rm eq}$ for $\epsilon=0$, and $\bar{B}_\ast = 0.3
\bar{B}_{\rm eq}$ for $\epsilon=1$. The magnitude $\bar{B}_\ast$
determines the level of the saturated mean magnetic field during
its nonlinear evolution (see Section V).

Let us determine the threshold for the generation of the mean
magnetic field due to the "shear-current" effect. To this end we
introduce the dynamo number in the kinematic approximation
\begin{eqnarray}
D = \biggl({l_0 \, L \, S \over \eta_{_{T}}} \biggr)^2 \,
\sigma_0(\bar{B}=0) \; .
\label{F10}
\end{eqnarray}
Consider the simple boundary conditions for a layer of the
thickness $2L$ in the $x$-direction: $B(|x|=1,z) = 0$ and
$A(|x|=1,z) = 0$, where $x$ is measured in units $L$. Then
Eqs.~(\ref{F11}) and (\ref{F12}) yield
\begin{eqnarray*}
B(t,x,z) &=& B_0 \exp(\gamma \, t) \, \cos(K_x \, x) \, \cos(K_z
\, z) \;,
\\
A(t,x,z) &=& - B_0 \, {l_0 \, \sqrt{\sigma_0} \over L} \,
\exp(\gamma \, t) \, \cos(K_x \, x) \, \sin(K_z \, z) \;,
\end{eqnarray*}
with the critical dynamo number $D_{\rm cr} = \pi^2$, where
$\sigma_0(\bar{B}=0) > 0$, the growth rate of the mean magnetic
field is $\gamma = \sqrt{D} \, K_z - K_x^2 - K_z^2 $, the wave
vector ${\bf K}$ is measured in units of $L^{-1}$ and the growth
rate $\gamma$ is measured in $ \eta_{_{T}} / L^{2} $. The mean
magnetic field is generated when $D > D_{\rm cr}$. The maximum
growth rate of the mean magnetic field, $ \gamma_{\rm max} = D^2 /
4 - K_x^2 ,$ is attained at $ K_z = K_m = \sqrt{D} / 2 .$ The
critical dynamo number determines the critical shear of the mean
velocity field $ S_{\rm cr} = (\pi / 3 \sqrt{\sigma_0}) (u_0 /
L)$. The scenario of a nonlinear evolution of the mean magnetic
field is discussed in Section V.

\section{Discussion}

In the present paper we developed the nonlinear theory of the
"shear-current" effect in a turbulence with an imposed mean
velocity shear. The ''shear-current" effect is associated with the
$\bar{\bf W} {\bf \times} \bar{\bf J}$-term in the mean
electromotive force and causes the generation of the mean magnetic
field even in a nonrotating and nonhelical homogeneous turbulence.
The scenario of the mean magnetic field evolution is as follows.
In the kinematic stage, the mean magnetic field grows due to the
''shear-current" effect from a very small seeding magnetic field.
During the nonlinear growth of the mean magnetic field, the
''shear-current" effect only changes its sign at some value
$\bar{B}_\ast$ of the mean magnetic field. However, there is no
quenching of the nonlinear "shear-current" effect contrary to the
quenching of the nonlinear $\alpha$-effect, the nonlinear
turbulent magnetic diffusion, etc. The magnitude $\bar{B}_\ast$ is
less than the equipartition field (see below). The background
magnetic fluctuations enhance the "shear-current" effect and
result in a decrease of the magnitude $\bar{\bf B}_\ast$.

\begin{figure}
\centering
\includegraphics[width=8cm]{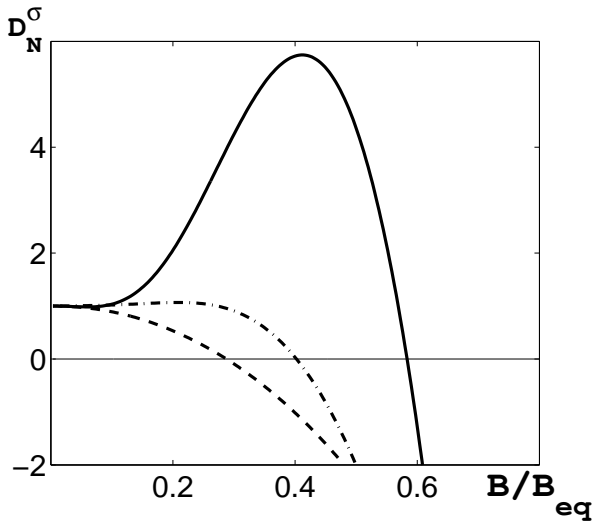}
\caption{\label{Fig4} The normalized nonlinear dynamo number
$D^\sigma_N(\bar{B})$ for different values of the parameter
$\epsilon$: $\, \, \, \epsilon=0$ (solid); $\epsilon = 0.2 $
(dashed-dotted); $\epsilon=1$ (dashed).}
\end{figure}

The magnitude $\bar{\bf B}_\ast$ determines the level of the
saturated mean magnetic field. Let us plot the normalized
nonlinear dynamo number $D^\sigma_N(\bar{B}) = D^\sigma(\bar{B}) /
D^\sigma(\bar{B}=0) $ which determines the role of the
''shear-current'' effect in the mean magnetic dynamo (see FIG. 4).
Here $D^\sigma(\bar{B}) = \sigma_0(\bar{B}) /
[\eta_{_{A}}(\bar{B}) \, \eta_{_{B}}(\bar{B})] $ is the nonlinear
dynamo number. At the point $\bar{B} = \bar{B}_\ast$ the nonlinear
effective dynamo number $D^\sigma_N(\bar{B}) = 0$. Depending on
the level of the background magnetic fluctuations described by the
parameter $\epsilon$, the saturated mean magnetic field varies
from $0.3 \bar{B}_{\rm eq}$ to $0.6 \bar{B}_{\rm eq}$ (see FIG.
4).

Note that the magnetic part of the $\alpha$ effect caused by the
magnetic helicity is a purely nonlinear effect. In this study we
concentrated on the algebraic nonlinearities (the nonlinear
"shear-current" effect, the nonlinear turbulent magnetic
diffusion, the nonlinear effective drift velocity of mean magnetic
field) and do not discuss the effect of magnetic helicity (the
dynamic nonlinearity, see, e.g.,
\cite{KR82,KRR94,GD94,KR99,KMRS2000,BB02}) on the nonlinear
saturation of the mean magnetic field. This is a subject of an
ongoing separate study. Note that the nonlinear "shear-current"
effect can affect the flux of magnetic helicity. However, this
remains an open issue.

The "shear-current" effect may be important in astrophysical
objects like accretion discs where mean velocity shear comes
together with rotation, so that both the "shear-current" effect
and the $\alpha$ effect might operate. Since the "shear-current"
effect is not quenched contrary to the quenching of the nonlinear
$\alpha$ effect, the "shear-current" effect might be the only
surviving effect, and it can explain the dynamics of large-scale
magnetic fields in astrophysical bodies with large-scale shearing
motions.

\appendix

\section{The nonlinear mean electromotive force in a turbulence
with a mean velocity shear}

We use a mean field approach whereby the velocity, pressure and
magnetic field are separated into the mean and fluctuating parts,
where the fluctuating parts have zero mean values. Let us derive
equations for the second moments. In order to exclude the pressure
term from the equation of motion~(\ref{B1}) we calculate $
\bec{\nabla} {\bf \times} (\bec{\nabla} {\bf \times} {\bf u}) .$
Then we rewrite the obtained equation and Eq.~(\ref{B2}) in a
Fourier space. We also apply the two-scale approach, e.g., a
correlation function
\begin{eqnarray*}
\langle u_i ({\bf x}) u_j ({\bf  y}) \rangle = \int \langle u_i
({\bf  k}_1) u_j ({\bf k}_2) \rangle  \exp \{i({\bf  k}_1 {\bf
\cdot} {\bf x}
\\
+ {\bf  k}_2 {\bf \cdot} {\bf y}) \} \,d{\bf k}_1 \, d{\bf  k}_2
\\
= \int f_{ij}({\bf k, K}) \exp{(i {\bf k} {\bf \cdot} {\bf r} + i
{\bf K} {\bf \cdot} {\bf R}) } \,d {\bf  k} \,d {\bf  K} \;,
\\
= \int f_{ij}({\bf k, R}) \exp{(i {\bf k} {\bf \cdot} {\bf r}) }
\,d {\bf  k} \;,
\end{eqnarray*}
where hereafter we omitted argument $t$ in the correlation
functions, $f_{ij}({\bf k, R}) = \hat L(u_i; u_j) ,$
\begin{eqnarray*}
\hat L(a; c) = \int \langle a({\bf k} + {\bf  K} / 2) c(-{\bf k} +
{\bf  K} / 2) \rangle
\\
\times \exp{(i {\bf K} {\bf \cdot} {\bf R}) } \,d {\bf  K} \;,
\end{eqnarray*}
and $ {\bf R} = ( {\bf x} +  {\bf y}) / 2  , \quad {\bf r} = {\bf
x} - {\bf y}, \quad {\bf K} = {\bf k}_1 + {\bf k}_2, \quad {\bf k}
= ({\bf k}_1 - {\bf k}_2) / 2 ,$ $ {\bf R} $ and $ {\bf K} $
correspond to the large scales, and $ {\bf r} $ and $ {\bf k} $ to
the  small ones (see, {\em e.g.,}  \cite{RS75,KR94}). This implies
that we assumed that there exists a separation of scales, i.e.,
the maximum scale of turbulent motions $ l_0 $ is much smaller
then the characteristic scale $L$ of inhomogeneities of the mean
fields. In particular, this implies that $ r \leq l_0 \ll R .$ Our
final results showed that this assumption is indeed valid. We
derive equations for the following correlation functions:
\begin{eqnarray*}
f_{ij}({\bf k, R}) &=& \hat L(u_i; u_j) \;, \; h_{ij}({\bf k, R})
= \hat L(b_i; b_j) \;,
\\
g_{ij}({\bf k, R}) &=& \hat L(b_i; u_j) \; .
\end{eqnarray*}
The equations for these correlation functions are given by
\begin{eqnarray}
{\partial f_{ij}({\bf k}) \over \partial t} &=& i({\bf k} {\bf
\cdot} \bar{\bf B}) \Phi_{ij}^{(M)} + I^f_{ij} +
I_{ijmn}^\sigma(\bar{\bf U}) f_{mn}
\nonumber \\
&& + F_{ij} + f_{ij}^N \;,
\label{B6} \\
{\partial h_{ij}({\bf k}) \over \partial t} &=& - i({\bf k}{\bf
\cdot} \bar{\bf B}) \Phi_{ij}^{(M)} + I^h_{ij} +
E_{ijmn}^\sigma(\bar{\bf U}) h_{mn} + h_{ij}^N \;,
\nonumber \\
\label{B7} \\
{\partial g_{ij}({\bf k }) \over \partial t} &=& i({\bf k} {\bf
\cdot} \bar{\bf B}) [f_{ij}({\bf k}) - h_{ij}({\bf k}) -
h_{ij}^{(H)}] + I^g_{ij}
\nonumber \\
&& + J_{ijmn}^\sigma(\bar{\bf U}) g_{mn} + g_{ij}^N \;,
\label{B8}
\end{eqnarray}
where hereafter we omitted argument $t$ and ${\bf R}$ in the
correlation functions and neglected terms $ \sim O(\nabla^2) .$
Here $\Phi_{ij}^{(M)}({\bf k}) = g_{ij}({\bf k}) - g_{ji}(-{\bf
k}) ,$ $ \, F_{ij}({\bf k}) = \langle \tilde F_i ({\bf k})
u_j(-{\bf k}) \rangle + \langle u_i({\bf k}) \tilde F_j(-{\bf k
})\rangle ,$ and ${\bf \tilde F} ({\bf k}) = {\bf k} {\bf \times}
({\bf k} {\bf \times} {\bf F}({\bf k})) / k^2 \rho .$ The tensors
$I_{ijmn}^\sigma(\bar{\bf U})$, $\, E_{ijmn}^\sigma(\bar{\bf U})$
and $J_{ijmn}^\sigma(\bar{\bf U})$ are given by
\begin{eqnarray*}
I_{ijmn}^\sigma(\bar{\bf U}) &=& \biggl[2 k_{iq} \delta_{mp}
\delta_{jn} + 2 k_{jq} \delta_{im} \delta_{pn} - \delta_{im}
\delta_{jq} \delta_{np}
\nonumber\\
&& - \delta_{iq} \delta_{jn} \delta_{mp} + \delta_{im} \delta_{jn}
k_{q} {\partial \over \partial k_{p}} \biggr] \nabla_{p} \bar
U_{q} \;,
\nonumber\\
E_{ijmn}^\sigma(\bar{\bf U}) &=& (\delta_{im} \delta_{jq} +
\delta_{jm} \delta_{iq}) \, \nabla_{n} \bar U_{q} \;,
\nonumber\\
J_{ijmn}^\sigma(\bar{\bf U}) &=& \biggl[2 k_{jq} \delta_{im}
\delta_{pn} - \delta_{im} \delta_{pn} \delta_{jq} + \delta_{jn}
\delta_{pm} \delta_{iq}
\nonumber\\
& & + \delta_{im} \delta_{jn} k_{q} {\partial \over \partial
k_{p}} \biggr] \nabla_{p} \bar U_{q} \;,
\end{eqnarray*}
where $\delta_{ij}$ is the Kronecker tensor, $ k_{ij} = k_i k_j /
k^2 $. Equation~(\ref{B6})-(\ref{B8}) are written in a frame
moving with a local velocity $ \bar {\bf U} $ of the mean flows.
In Eqs.~(\ref{B6}) and (\ref{B8}) we neglected small terms which
are of the order of $O(|\nabla^2 \bar{\bf U}|) .$ The source terms
$I_{ij}^f$ , $\, I_{ij}^h$ and $I_{ij}^g$ which contain the
large-scale spatial derivatives of the mean magnetic field are
given by
\begin{eqnarray}
I_{ij}^f &=& {1 \over 2}(\bar{\bf B} {\bf \cdot} \bec{\nabla})
\Phi_{ij}^{(P)} + [g_{qj}({\bf k}) (2 P_{in}(k) - \delta_{in})
\nonumber \\
&& + g_{qi}(-{\bf k}) (2 P_{jn}(k) - \delta_{jn})] \bar{B}_{n,q} -
\bar{B}_{n,q} k_{n} \Phi_{ijq}^{(P)}\;,
\nonumber\\
\label{M1}\\
I_{ij}^h &=& {1 \over 2}(\bar{\bf B} {\bf \cdot} \bec{\nabla})
\Phi_{ij}^{(P)} - [g_{iq}({\bf k}) \delta_{jn} + g_{jq}(-{\bf k})
\delta_{in}] \bar{B}_{n,q}
\nonumber\\
&& - \bar{B}_{n,q} k_{n} \Phi_{ijq}^{(P)} \;,
\label{M2}\\
I_{ij}^g &=& {1 \over 2} (\bar{\bf B} {\bf \cdot} \bec{\nabla})
(f_{ij} + h_{ij}) + h_{iq} (2 P_{jn}(k) - \delta_{jn})
\bar{B}_{n,q}
\nonumber \\
&& - f_{nj} \bar{B}_{i,n} - \bar{B}_{n,q} k_{n}(f_{ijq} + h_{ijq})
\;,
\label{M3}
\end{eqnarray}
where $ \bec{\nabla} = \partial / \partial {\bf R} $, $\,
\Phi_{ij}^{(P)}({\bf k}) = g_{ij}({\bf k}) + g_{ji}(-{\bf k}) ,$
and $\bar{B}_{i,j} = \nabla_j \bar{B}_{i} ,$ the terms $ \,
f_{ij}^{N} ,$ $\, h_{ij}^{N} $ and $ g_{ij}^{N} $ are determined
by the third moments appearing due to the nonlinear terms,
$f_{ijq} = (1/2) \partial f_{ij} / \partial k_{q} ,$ and similarly
for $h_{ijq}$ and $\Phi_{ijq}^{(P)}$. A stirring force in the
Navier-Stokes turbulence is an external parameter, that determines
the background turbulence.

For the derivation of Eqs. (\ref{B6})-(\ref{B8}) we performed
several calculations that are similar to the following, which
arose, e.g., in computing $ \partial g_{ij} / \partial t .$ The
other calculations follow similar lines and are not given here.
Let us define $ Y_{ij}({\bf k, R}) $ by
\begin{eqnarray}
&& Y_{ij}({\bf k, R}) = i \int (k_{p} + K_{p}/2) \bar B_{p}({\bf
Q}) \exp(i {\bf K} {\bf \cdot} {\bf R})
\nonumber\\
&& \times \langle u_i ({\bf k} + {\bf  K} / 2 - {\bf  Q})
u_j(-{\bf k} + {\bf  K}  / 2) \rangle \,d {\bf  K} \,d {\bf  Q}
\; . \label{P23}
\end{eqnarray}
Next, we introduce new variables:
\begin{eqnarray}
\tilde {\bf k}_{1} &=& {\bf k} + {\bf  K} / 2 - {\bf  Q} \;, \quad
\tilde {\bf k}_{2} = - {\bf k} + {\bf  K} / 2 \;, \label{P24}
\\
\tilde {\bf k} &=& (\tilde {\bf k}_{1} - \tilde {\bf k}_{2}) / 2 =
{\bf k} - {\bf  Q} / 2 \;, \; \tilde {\bf K} = \tilde {\bf k}_{1}
+ \tilde {\bf k}_{2} = {\bf  K} - {\bf  Q} . \nonumber
\end{eqnarray}
Therefore,
\begin{eqnarray}
Y_{ij}({\bf k, R} ) &=& i \int f_{ij}({\bf k} - {\bf  Q} / 2, {\bf
K} - {\bf  Q}) (k_{p} + K_{p}/2) \bar B_{p}({\bf  Q})
\nonumber\\
& & \times \exp{(i {\bf K} {\bf \cdot} {\bf R})} \,d {\bf  K} \,d
{\bf  Q} \; .
\label{B12}
\end{eqnarray}
Since $ |{\bf Q}| \ll |{\bf k}| $ we use the Taylor expansion
\begin{eqnarray}
f_{ij}({\bf k} - {\bf Q}/2, {\bf  K} - {\bf  Q}) \simeq
f_{ij}({\bf k},{\bf  K} - {\bf  Q})
\nonumber\\
- \frac{1}{2} {\partial f_{ij}({\bf k},{\bf  K} - {\bf Q}) \over
\partial k_s} Q_s  + O({\bf Q}^2) \;,
\label{B14}
\end{eqnarray}
and the following identities:
\begin{eqnarray}
&& [f_{ij}({\bf k},{\bf R}) \bar B_{p}({\bf R})]_{\bf  K} = \int
f_{ij}({\bf k},{\bf  K} - {\bf  Q}) \bar B_{p}({\bf Q}) \,d {\bf
Q} \;,
\nonumber \\
&& \nabla_{p} [f_{ij}({\bf k},{\bf R}) \bar B_{p}({\bf R})] = \int
i K_{p} [f_{ij}({\bf k},{\bf R}) \bar B_{p}({\bf R})]_{\bf  K}
\nonumber\\
&& \times \exp{(i {\bf K} {\bf \cdot} {\bf R})} \,d {\bf  K} \; .
\label{B15}
\end{eqnarray}
Therefore, Eqs. (\ref{B12})-(\ref{B15}) yield
\begin{eqnarray}
Y_{ij}({\bf k},{\bf R}) &\simeq& [i({\bf k } \cdot \bar{\bf B}) +
(1/2) (\bar{\bf B} \cdot \bec{\nabla})] f_{ij}({\bf k},{\bf R})
\nonumber\\
& & - \frac{1}{2} k_{p} {\partial f_{ij}({\bf k}) \over
\partial k_s} \bar B_{p,s} \; .
\label{B16}
\end{eqnarray}
We took into account that in Eq. (\ref{B8}) the terms with
symmetric tensors with respect to the indexes "i" and "j" do not
contribute to the mean electromotive force because ${\cal E}_{m} =
\varepsilon_{mji} \, g_{ij} $. In Eqs. (\ref{B12})-(\ref{B15}) we
neglected the second and higher derivatives over $ {\bf R} .$ For
the derivation of Eqs. (\ref{B6})-(\ref{B8}) we also used the
following identity
\begin{eqnarray}
&& i k_i \int f_{ij}({\bf k} - {1 \over 2}{\bf  Q}, {\bf K} - {\bf
Q}) \bar U_{p}({\bf  Q}) \exp(i {\bf K} {\bf \cdot} {\bf R}) \,d
{\bf  K} \,d {\bf  Q}
\nonumber\\
& & = -{1 \over 2} \bar U_{p} \nabla _i f_{ij} + {1 \over 2}
f_{ij} \nabla _i \bar U_{p} -  {i \over 4} (\nabla _s \bar U_{p})
\biggl(\nabla _i {\partial f_{ij} \over \partial k_s} \biggr)
\nonumber\\
& & +  {i \over 4} \biggl( {\partial f_{ij} \over
\partial k_s} \biggr) (\nabla _s \nabla _i \bar U_{p}) \; .
\label{D100}
\end{eqnarray}
To derive Eq.~(\ref{D100}) we multiply  the equation
$\bec{\nabla}~\cdot~{\bf u} = 0$ [written in ${\bf k}$-space for
$u_i({\bf k}_1 - {\bf Q})]$ by $u_j({\bf k}_2) \bar U_{p}({\bf Q})
\exp(i {\bf K} {\bf \cdot} {\bf R}) ,$ integrate over ${\bf K}$
and ${\bf Q}$, and average over ensemble of velocity fluctuations.
Here ${\bf k}_1 = {\bf k} + {\bf  K} / 2$ and ${\bf k}_2 = -{\bf
k} + {\bf K} / 2 .$ This yields
\begin{eqnarray}
&& \int i \biggl(k_i + {1 \over 2} K_i - Q_i \biggr) \langle
u_i({\bf k} + {1 \over 2}{\bf K} - {\bf Q}) u_j(-{\bf k} + {1
\over 2}{\bf K}) \rangle
\nonumber\\
& & \times \bar U_{p}({\bf  Q}) \exp{(i {\bf K} {\bf \cdot} {\bf
R})} \,d {\bf  K} \,d {\bf Q} = 0  \; . \label{D2}
\end{eqnarray}
Now we introduce new variables, $ \tilde {\bf k}_{1}$ and $\tilde
{\bf k}_{2}$ determined by Eq.~(\ref{P24}). This allows us to
rewrite Eq.~(\ref{D2}) in the form
\begin{eqnarray}
& & \int i \biggl(k_i + {1 \over 2} K_i - Q_i \biggr) f_{ij}({\bf
k} - {1 \over 2}{\bf Q}, {\bf K} - {\bf Q}) \bar U_{p}({\bf  Q})
\nonumber\\
& & \times \exp{(i {\bf K} {\bf \cdot} {\bf R})} \,d {\bf  K} \,d
{\bf Q} = 0  \; . \label{D3}
\end{eqnarray}
Since $ |{\bf Q}| \ll |{\bf k}| $ we use the Taylor
expansion~(\ref{B14}), and we also use the following identities,
which are similar to Eq.~(\ref{B15}):
\begin{eqnarray}
&& [f_{ij}({\bf k},{\bf R}) \bar U_{p}({\bf R})]_{\bf  K} = \int
f_{ij}({\bf k},{\bf  K} - {\bf  Q}) \bar U_{p}({\bf Q}) \,d {\bf
Q} \;,
\nonumber \\
&& \nabla_{p} [f_{ij}({\bf k},{\bf R}) \bar U_{p}({\bf R})] = \int
i K_{p} [f_{ij}({\bf k},{\bf R}) \bar U_{p}({\bf R})]_{\bf  K}
\nonumber\\
&& \times \exp{(i {\bf K} {\bf \cdot} {\bf R})} \,d {\bf  K} \; .
\label{D5}
\end{eqnarray}
Therefore, Eq.~(\ref{D3}) yields Eq.~(\ref{D100}).

Now we split all tensors into nonhelical, $f_{ij},$ and helical,
$f_{ij}^{(H)},$ parts. Note that the helical part of the tensor of
magnetic fluctuations $h_{ij}^{(H)}$ depends on the magnetic
helicity and it is not determined by Eq.~(\ref{B7}). The equation
for the helical part of the tensor of magnetic fluctuations
$h_{ij}^{(H)}$ follows from the magnetic helicity conservation
arguments (see, {\rm e.g.,}
\cite{KR82,KRR94,GD94,KR99,KMRS2000,BB02}).

In this study we use the $\tau$ approximation [see Eq.
(\ref{A1})]. The $ \tau $-approximation  is in general similar to
Eddy Damped Quasi Normal Markovian (EDQNM) approximation. However
some principle difference exists between these two approaches (see
\cite{O70,Mc90}). The EDQNM closures do not relax to equilibrium,
and this procedure does not describe properly the motions in the
equilibrium state in contrast to the $ \tau $-approximation.
Within the EDQNM theory, there is no dynamically determined
relaxation time, and no slightly perturbed steady state can be
approached \cite{O70}. In the $ \tau $-approximation, the
relaxation time for small departures from equilibrium is
determined by the random motions in the equilibrium state, but not
by the departure from equilibrium \cite{O70}. As follows from the
analysis by \cite{O70} the $ \tau $-approximation describes the
relaxation to equilibrium state (the background turbulence) much
more accurately than the EDQNM approach.

\subsection{Shear-free homogeneous turbulence}

Consider a turbulence without a mean velocity shear, i.e., we omit
tensors $I_{ijmn}^\sigma(\bar{\bf U})$, $\,
E_{ijmn}^\sigma(\bar{\bf U})$ and $J_{ijmn}^\sigma(\bar{\bf U})$
in Eqs.~(\ref{B6})-(\ref{B8}). First we solve
Eqs.~(\ref{B6})-(\ref{B8}) neglecting the sources $I^f_{ij},
I^h_{ij}, I^g_{ij}$ with the large-scale spatial derivatives. Then
we will take into account the terms with the large-scale spatial
derivatives by perturbations. We start with
Eqs.~(\ref{B6})-(\ref{B8}) written for nonhelical parts of the
tensors, and then consider Eqs.~(\ref{B6})-(\ref{B8}) for helical
parts of the tensors.

Thus, we subtract Eqs. (\ref{B6})-(\ref{B8}) written for
background turbulence (for $\bar{\bf B}=0)$ from those for
$\bar{\bf B} \not=0$. Then we use the $\tau$ approximation and
neglect the terms with the large-scale spatial derivatives. Next,
we assume that $\eta k^2 \ll \tau^{-1}$ and $\nu k^2 \ll
\tau^{-1}$ for the inertial range of turbulent flow, and we also
assume that the characteristic time of variation of the mean
magnetic field $\bar{\bf B}$ is substantially larger than the
correlation time $\tau(k)$ for all turbulence scales. Thus, we
arrive to the following steady-state solution of the obtained
equations:
\begin{eqnarray}
\hat f_{ij}({\bf k}) &\approx& f_{ij}^{(0)}({\bf k}) + i \tau
({\bf k} {\bf \cdot} \bar{\bf B}) \hat \Phi_{ij}^{(M)}({\bf k})
\;,
\label{B17}\\
\hat h_{ij}({\bf k}) &\approx& h_{ij}^{(0)}({\bf k}) - i \tau
({\bf k} {\bf \cdot} \bar{\bf B}) \hat \Phi_{ij}^{(M)}({\bf k})
\;,
\label{B18}\\
\hat g_{ij}({\bf k}) &\approx& i \tau ({\bf k} {\bf \cdot}
\bar{\bf B}) [\hat f_{ij}({\bf k}) - \hat h_{ij}({\bf k})] \;,
\label{B19}
\end{eqnarray}
where $\hat f_{ij}, \hat h_{ij}$ and $\hat g_{ij}$ are solutions
without the sources $I^f_{ij}, I^h_{ij}$ and $I^g_{ij}$.

Now we split all correlation functions into symmetric and
antisymmetric parts with respect to the wave number ${\bf k}$,
{\em e.g.,} $ f_{ij} = f_{ij}^{(s)} + f_{ij}^{(a)} ,$ where $
f_{ij}^{(s)} = [f_{ij}({\bf k}) + f_{ij}(-{\bf k})] / 2 $ is the
symmetric part and $ f_{ij}^{(a)} = [f_{ij}({\bf k}) -
f_{ij}(-{\bf k})] / 2 $ is the antisymmetric part, and similarly
for other tensors. Thus, Eqs. (\ref{B17})-(\ref{B19}) yield
\begin{eqnarray}
\hat f_{ij}^{(s)}({\bf k}) &\approx& {1 \over 1 + 2 \psi} [(1 +
\psi) f_{ij}^{(0)}({\bf k}) + \psi h_{ij}^{(0)}({\bf k})]  \;,
\nonumber\\
\label{B22} \\
\hat h_{ij}^{(s)}({\bf k}) &\approx& {1 \over 1 + 2 \psi} [\psi
f_{ij}^{(0)}({\bf k}) + (1 + \psi) h_{ij}^{(0)}({\bf k})] \;,
\nonumber\\
\label{B24}\\
\hat g_{ij}^{(a)}({\bf k}) &\approx& {i \tau ({\bf k} {\bf \cdot}
\bar{\bf B}) \over 1 + 2 \psi} [f_{ij}^{(0)}({\bf k}) -
h_{ij}^{(0)}({\bf k})]  \;, \label{B26}
\end{eqnarray}
where $ \psi({\bf k}) = 2 (\tau \, {\bf k} {\bf \cdot} \bar{\bf
B})^2 .$ The correlation functions $\hat f_{ij}^{(a)}({\bf k})$,
$\hat h_{ij}^{(a)}({\bf k})$ and $\hat g_{ij}^{(s)}({\bf k})$
vanish if we neglect the large-scale spatial derivatives, i.e.,
they are proportional to the first-order spatial derivatives.
Equations~(\ref{B22}) and (\ref{B24}) yield
\begin{eqnarray}
\hat f_{ij}^{(s)}({\bf k}) + \hat h_{ij}^{(s)}({\bf k}) \approx
f_{ij}^{(0)}({\bf k}) + h_{ij}^{(0)}({\bf k}) \;, \label{B27}
\end{eqnarray}
which is in agreement with that a uniform mean magnetic field
performs no work on the turbulence. A uniform mean magnetic field
can only redistribute the energy between hydrodynamic fluctuations
and magnetic fluctuation. A change of the total energy of
fluctuations is caused by a nonuniform mean magnetic field.

Next, we take into account the large-scale spatial derivatives in
Eqs. (\ref{B6})-(\ref{B8}) by perturbations. Their effect
determines the following steady-state equations for the second
moments $\tilde f_{ij}$, $\tilde h_{ij}$ and $\tilde g_{ij}$:
\begin{eqnarray}
\tilde f_{ij}^{(a)}({\bf k}) &=& i \tau ({\bf k} {\bf \cdot}
\bar{\bf B}) \tilde \Phi_{ij}^{(M,s)}({\bf k}) + \tau I^f_{ij} \;,
\label{B28}\\
\tilde h_{ij}^{(a)}({\bf k}) &=& - i \tau ({\bf k} {\bf \cdot}
\bar{\bf B}) \tilde \Phi_{ij}^{(M,s)}({\bf k}) + \tau I^h_{ij} \;,
\label{B29} \\
\tilde g_{ij}^{(s)}({\bf k }) &=& i \tau ({\bf k} {\bf \cdot}
\bar{\bf B}) (\tilde f_{ij}^{(a)}({\bf k}) - \tilde
h_{ij}^{(a)}({\bf k})) + \tau I^g_{ij} \;,
\nonumber \\
\label{B30}
\end{eqnarray}
where $ \tilde \Phi_{ij}^{(M,s)} = [\tilde \Phi_{ij}^{(M)}({\bf
k}) + \tilde \Phi_{ij}^{(M)}(-{\bf k})] / 2 .$ The solution of
Eqs. (\ref{B28})-(\ref{B30}) yield
\begin{eqnarray}
\tilde \Phi_{ij}^{(M,s)}({\bf k}) &=& {\tau \over 1 + 2 \psi} \{
I^g_{ij} - I^g_{ji} + i \tau ({\bf k} {\bf \cdot} \bar{\bf B})
(I^f_{ij} - I^f_{ji}
\nonumber\\
& & + I^h_{ji} - I^h_{ij}) \} \; . \label{B31}
\end{eqnarray}
Substituting Eq.~(\ref{B31}) into Eqs. (\ref{B28})-(\ref{B30}) we
obtain the final expressions in ${\bf k}$-space for the nonhelical
parts of the tensors $\tilde f_{ij}^{(a)}({\bf k})$, $\tilde
h_{ij}^{(a)}({\bf k})$, $\tilde g_{ij}^{(s)}({\bf k})$ and $\tilde
\Phi_{ij}^{(M,s)}({\bf k})$. In particular,
\begin{eqnarray}
&& \tilde \Phi_{mn}^{(M,s)}({\bf k}) = {\tau \over (1 + 2 \psi)^2}
\biggl[(1 + \epsilon)(1 + 2 \psi) (\delta_{nj} \delta_{mk}
\nonumber\\
& & - \delta_{mj} \delta_{nk} + k_{nk} \delta_{mj} - k_{mk}
\delta_{nj}) - 2 (\epsilon + 2 \psi) (k_{nj} \delta_{mk}
\nonumber\\
& & - k_{mj} \delta_{nk}) \biggr] \, \bar B_{j,k} \; .
\label{BB40}
\end{eqnarray}

The correlation functions $\tilde f_{ij}^{(s)}({\bf k})$, $\tilde
h_{ij}^{(s)}({\bf k})$ and $\tilde g_{ij}^{(a)}({\bf k})$ are of
the order of $\sim O(\nabla^2)$, i.e., they are proportional to
the second-order spatial derivatives. Thus $\hat f_{ij} + \tilde
f_{ij} $ is the nonhelical part of the correlation functions for a
shear-free turbulence, and similarly for other second moments.

Now we solve Eqs.~(\ref{B6})-(\ref{B8}) for helical parts of the
tensors for a shear-free turbulence using the same approach which
we used in this section. The steady-state solution of
Eqs.~(\ref{B6}) and (\ref{B8}) for the helical parts of the
tensors reads:
\begin{eqnarray}
f_{ij}^{(H)}({\bf k}) &\approx& i \tau ({\bf k} {\bf \cdot}
\bar{\bf B}) \Phi_{ij}^{(M,H)}({\bf k}) \;,
\label{B32}\\
g_{ij}^{(H)}({\bf k}) &\approx& i \tau ({\bf k} {\bf \cdot}
\bar{\bf B}) [f_{ij}^{(H)}({\bf k}) - h_{ij}^{(H)}({\bf k})] \;,
\label{B33}
\end{eqnarray}
where $\Phi_{ij}^{(M,H)}({\bf k}) = g_{ij}^{(H)}({\bf k}) -
g_{ji}^{(H)}(-{\bf k}) $. The tensor $h_{ij}^{(H)}$ is determined
by the dynamic equation (see, {\rm e.g.,}
\cite{KR82,KRR94,GD94,KR99,KMRS2000,BB02}). The solution of Eqs.
(\ref{B32}) and (\ref{B33}) yield
\begin{eqnarray}
\Phi_{ij}^{(M,H)}({\bf k}) &=& - {2 i \tau ({\bf k} {\bf \cdot}
\bar{\bf B}) \over 1 + \psi} h_{ij}^{(H)} \; . \label{B34}
\end{eqnarray}
Since $h_{ij}^{(H)}$ is of the order of $O(\nabla)$ we do not need
to take into account the source terms with the large-scale spatial
derivatives \cite{KR99}.

Now we calculate the mean electromotive force $ {\cal E}_{i}({\bf
r}=0) = (1/2) \varepsilon_{inm} \int [\Phi_{mn}^{(M,H)}({\bf k}) +
\tilde \Phi_{mn}^{(M,s)}({\bf k})] \,d {\bf k} $. Thus,
\begin{eqnarray}
{\cal E}_{i} &=& \varepsilon_{inm} \int \biggl[{\tau \over 1 + 2
\psi} \, [I^g_{mn} + i \tau ({\bf k} {\bf \cdot} \bar{\bf B})
(I^f_{mn} - I^h_{mn})]
\nonumber\\
& & - {i \tau ({\bf k} {\bf \cdot} \bar{\bf B}) \over 1 + \psi}
h_{mn}^{(H)} \biggr] \,d {\bf k} \; . \label{B35}
\end{eqnarray}
For the integration in $ {\bf k} $-space of the mean electromotive
force we have to specify a model for the background turbulence
(with $ \bar{\bf B} = 0),$ see Section III. After the integration
in $ {\bf k} $-space we obtain ${\cal E}_{i} = a_{ij} \bar B_{j} +
b_{ijk} \bar B_{j,k} ,$ where
\begin{eqnarray}
a_{ij} &=& - i \varepsilon_{inm} \int {\tau k_j h_{mn}^{(H)} \over
1 + \psi} \,d {\bf k} = \chi^{(c)}(\bar{\bf B})
\biggl[\phi^{(m)}(\beta) \beta_{ij}
\nonumber\\
& & + {1 \over 2} \biggl(3 - (1 + \beta^{2}) \phi^{(m)}(\beta)
\biggr) P_{ij}(\beta) \biggr] \;,
\label{B36}\\
b_{ijk} &=& {1 \over 2} \eta_{_{T}} \biggl[ (1 + \epsilon)
\varepsilon_{ijm} \biggl(\delta_{km} K_{pp}^{(1)}(\sqrt{2} \beta)
- K_{km}^{(1)}(\sqrt{2} \beta) \biggr)
\nonumber\\
& & + 2 \varepsilon_{ink} \biggl((1 - \epsilon) \tilde
\Psi_1\{K_{jn}\} - K_{jn}^{(1)}(\sqrt{2} \beta)\biggr) \biggr] \;,
\nonumber\\
\label{B37}
\end{eqnarray}
\begin{eqnarray*}
\tilde \Psi_1\{X\} &=& 3 X^{(1)}(\sqrt{2} \beta) - {3 \over 2 \pi}
\bar X(2 \beta^2) \;,
\\
K_{ij}^{(1)}(\beta) &=& {3 \over 2 \pi} \int_0^1 \bar
K_{ij}(a(\bar \tau)) \bar \tau \, d\bar \tau
\nonumber\\
& & = {3 \beta^{4} \over \pi} \int_{\beta}^{\infty} {\bar
K_{ij}(X^{2}) \over X^{5}} \,d X \;,
\end{eqnarray*}
and all calculations are made for $q=5/3$, $\, X^{2} = \beta^{2}
(k / k_{0})^{2/3} = a = [\beta u_{0} k \tau(k) / 2]^{2}$, the
function $\bar K_{ij}$ is determined by Eq.~(\ref{C22}) in
Appendix B, and $\, \epsilon = \langle {\bf b}^2 \rangle^{(0)} /
\langle {\bf u}^2 \rangle^{(0)}$, $ \, \beta = 4 \bar B / (u_{0}
\sqrt{2 \mu \rho}) ,$ $\, P_{ij}(\beta) = \delta_{ij} -
\beta_{ij}$, $\beta_{ij} = \bar B_{i} \bar B_{j} / \bar B^2$, $\,
\phi^{(m)}(\beta) = (3 / \beta^{2}) (1 - \arctan (\beta) / \beta)
,$ $\, \chi^{(c)}(\bar {\bf B}) \equiv (\tau / 3 \mu \rho) \langle
{\bf b} \cdot (\bec{\nabla} {\bf \times} {\bf b}) \rangle$ is
related with current helicity. Since a part of the mean
electromotive force is determined by the function $
a_{ij}(\bar{\bf B}) \bar B_{j} $ and $ P_{ij}(\beta) \bar B_{j} =
0$, we can drop the term $ \propto P_{ij}(\beta) $ in
Eq.~(\ref{B36}). Thus, the equations for $a_{ij}$ and $b_{ijk}$
are given by
\begin{eqnarray}
a_{ij} &=& \alpha^{(m)}(\bar{\bf B}) \, \delta_{ij} \;,
\label{L1}\\
b_{ijk} &=& \eta_{_{T}} \biggl[b_1 \, \varepsilon_{ijk} + b_2 \,
\varepsilon_{ijn} \, \beta_{nk} + b_3 \, \varepsilon_{ink} \,
\beta_{nj} \biggr] \;,
\nonumber\\
\label{L2}
\end{eqnarray}
where $\alpha^{(m)}(\bar{\bf B}) = \chi^{(c)}(\bar{\bf B}) \,
\phi^{(m)}(\beta)$, and
\begin{eqnarray*}
b_1 &=& A_{1}^{(1)}(\sqrt{2} \beta) + A_{2}^{(1)}(\sqrt{2} \beta)
\;,
\\
b_2 &=& - {1 \over 2} (1 + \epsilon) A_{2}^{(1)}(\sqrt{2} \beta)
\;,
\\
b_3 &=& (1 - \epsilon) \tilde \Psi_1\{A_{2}\} -
A_{2}^{(1)}(\sqrt{2} \beta) \;,
\end{eqnarray*}
the functions $\bar A_{k}(y)$ and $A_{k}^{(1)}(y)$ are determined
by Eqs.~(\ref{P22}) and ~(\ref{P20}) in Appendixes B and C.
Equations~(\ref{L1}) and~(\ref{L2}) yield
Eqs.~(\ref{L20})-(\ref{B41}).

\subsection{Turbulence with a mean velocity shear}

Now we study the effect of a mean velocity shear on the mean
electromotive force. We take into account the tensors
$I_{ijmn}^\sigma(\bar{\bf U})$, $\, E_{ijmn}^\sigma(\bar{\bf U})$
and $J_{ijmn}^\sigma(\bar{\bf U})$ in Eqs.~(\ref{B6})-(\ref{B8}),
and we neglect terms $ \sim O(\nabla^2) .$ The steady-state
solution of Eqs.~(\ref{B6})-(\ref{B8}) for the nonhelical parts of
the tensors for a sheared turbulence reads:
\begin{eqnarray}
N^f_{ijmn}(\bar{\bf U}) f_{mn} &=& \tau \{ i({\bf k} {\bf \cdot}
\bar{\bf B}) \Phi_{ij}^{(M)} + I^f_{ij} \} \;,
\label{S1} \\
N^h_{ijmn}(\bar{\bf U}) h_{mn} &=& \tau \{-i({\bf k} {\bf \cdot}
\bar{\bf B}) \Phi_{ij}^{(M)} + I^h_{ij} \} \;,
\label{S100}\\
N^g_{ijmn}(\bar{\bf U}) g_{mn} &=& \tau \{ i({\bf k} {\bf \cdot}
\bar{\bf B}) [f_{ij}({\bf k}) - h_{ij}({\bf k})] + I^g_{ij} \} \;,
\nonumber \\
\label{S2}
\end{eqnarray}
where
\begin{eqnarray*}
N^f_{ijmn}(\bar{\bf U}) &=& \delta_{im} \delta_{jn} - \tau
I_{ijmn}^\sigma \;,
\\
N^h_{ijmn}(\bar{\bf U}) &=& \delta_{im} \delta_{jn} - \tau
E_{ijmn}^\sigma \;,
\\
N^g_{ijmn}(\bar{\bf U}) &=& \delta_{im} \delta_{jn} - \tau
J_{ijmn}^\sigma \;,
\end{eqnarray*}
and we use the following notations: the total correlation function
is $f_{ij} = \bar f_{ij} + f_{ij}^{\sigma}$. Here $ \bar f_{ij} =
\hat f_{ij} + \tilde f_{ij} $ is the correlation functions for a
shear-free turbulence, and the correlation functions
$f_{ij}^{\sigma}$ determines the effect of a mean velocity shear.
The similar notations are for other correlation functions. Now we
solve Eqs.~(\ref{S1})-(\ref{S2}) by iterations. This yields
\begin{eqnarray}
f_{ij}^{\sigma}({\bf k}) &=& \tau \{I_{ijmn}^\sigma \bar f_{mn} +
i({\bf k} {\bf \cdot} \bar{\bf B}) \Phi_{ij}^{(M,\sigma)} +
I^{(f,\sigma)}_{ij} \} \;,
\nonumber\\
\label{S3} \\
h_{ij}^{\sigma}({\bf k}) &=& \tau \{E_{ijmn}^\sigma \bar h_{mn} -
i({\bf k} {\bf \cdot} \bar{\bf B}) \Phi_{ij}^{(M,\sigma)} +
I^{(h,\sigma)}_{ij} \} \;,
\nonumber\\
\label{S4} \\
g_{ij}^{\sigma}({\bf k}) &=& \tau \{J_{ijmn}^\sigma \bar g_{mn} +
i ({\bf k} {\bf \cdot} \bar{\bf B}) [f_{ij}^{\sigma} -
h_{ij}^{\sigma}] + I^{(g,\sigma)}_{ij} \} \;,
\nonumber\\
\label{S5}
\end{eqnarray}
where $\Phi_{ij}^{(M,\sigma)}({\bf k}) = g_{ij}^{\sigma}({\bf k})
- g_{ji}^{\sigma}(-{\bf k}) $, the source terms
$I^{(f,\sigma)}_{ij} \equiv I^{f}_{ij}(g_{ij}^{\sigma})$, $\,
I^{(h,\sigma)}_{ij} \equiv I^{h}_{ij}(g_{ij}^{\sigma})$ and
$I^{(g,\sigma)}_{ij} \equiv I^{g}_{ij}(f_{ij}^{\sigma},
h_{ij}^{\sigma})$ are determined by Eqs.~(\ref{M1})-(\ref{M3}),
where $f_{ij}$, $h_{ij}$, $g_{ij}$ are replaced by
$f_{ij}^{\sigma}$, $h_{ij}^{\sigma}$, $g_{ij}^{\sigma}$,
respectively. The solution of Eqs. (\ref{S3})-(\ref{S5}) yield
equation for the symmetric part $\Phi_{ij}^{(M,\sigma,s)}$ of the
tensor:
\begin{eqnarray}
&& \Phi_{ij}^{(M,\sigma,s)}({\bf k}) = {\tau \over 1 + 2 \psi} \{
(J_{ijmn}^\sigma - J_{jimn}^\sigma) \tilde g_{mn} +
I^{(g,\sigma)}_{ij}
\nonumber\\
& & - I^{(g,\sigma)}_{ji} + i \tau ({\bf k} {\bf \cdot} \bar{\bf
B}) [(I_{ijmn}^\sigma - I_{jimn}^\sigma) \tilde f_{mn} +
I^{(f,\sigma)}_{ij}
\nonumber\\
& &  - I^{(f,\sigma)}_{ji} + I^{(h,\sigma)}_{ji} -
I^{(h,\sigma)}_{ij}] \} \;, \label{S6}
\end{eqnarray}
where we took into account that $E_{ijmn}^\sigma$ is a symmetric
tensor in indexes $i$ and $j$. Thus, the effect of a mean velocity
shear on the mean electromotive force, $ {\cal E}^\sigma_{i}({\bf
r}=0) \equiv (1/2) \varepsilon_{inm} \int \Phi_{mn}^{(M,\sigma,s)}
\,d {\bf k} ,$ is determined by
\begin{eqnarray}
{\cal E}^\sigma_{i} &=& \varepsilon_{inm} \int {\tau \over 1 + 2
\psi} \{J_{mnpq}^\sigma \tilde g_{pq} + i \tau ({\bf k} {\bf
\cdot} \bar{\bf B}) [I_{mnpq}^\sigma \tilde f_{pq}
\nonumber \\
& & + I^{(f,\sigma)}_{mn} - I^{(h,\sigma)}_{mn}] +
I^{(g,\sigma)}_{mn}\} \, d{\bf k} \; .
\label{S7}
\end{eqnarray}

Now we use the following identities:
\begin{widetext}
\begin{eqnarray}
\varepsilon_{imp} \bar K_{jkqm} \nabla_p \bar U_q &=& {1 \over 2}
[\bar C_1 (2 S_{ijk}^{(1)} + 2 S_{ijk}^{(2)} + S_{ijk}^{(4)} +
S_{ijk}^{(5)}) + \bar C_3 (2 S_{ijk}^{(6)} + S_{ijk}^{(7)})] \;,
\quad \varepsilon_{ijm} \bar K_{kmpq} \nabla_p \bar U_q = 2 \bar
C_1 S_{ijk}^{(1)} \;,
\nonumber\\
\varepsilon_{ikp} \bar K_{jq} \nabla_p \bar U_q  &=& {1 \over 2}
[\bar A_1 (2 S_{ijk}^{(2)} + S_{ijk}^{(4)}) + \bar A_2 (2
S_{ijk}^{(6)} + S_{ijk}^{(7)})] \;, \quad \varepsilon_{imk} \bar
K_{jmpq} \nabla_p \bar U_q = -2 (\bar C_1 S_{ijk}^{(2)} + \bar C_3
S_{ijk}^{(6)}) \;,
\nonumber\\
\varepsilon_{imq} \bar K_{mj} \nabla_k \bar U_q  &=&
\varepsilon_{ijm} \bar K_{mq} \nabla_k \bar U_q = {1 \over 2} \bar
A_1 (2 S_{ijk}^{(1)} - S_{ijk}^{(5)}) \;, \quad \varepsilon_{ikq}
\bar K_{mm} \nabla_j \bar U_q  = {1 \over 2} (3 \bar A_1 + \bar
A_2) (2 S_{ijk}^{(2)} - S_{ijk}^{(4)}) \;,
\nonumber\\
\varepsilon_{imk} \bar K_{mq} \nabla_j \bar U_q &=& -
\varepsilon_{imq} \bar K_{mk} \nabla_j \bar U_q = - {1 \over 2}
\bar A_1 (2 S_{ijk}^{(2)} - S_{ijk}^{(4)}) \;, \quad
\varepsilon_{ijp} \bar K_{kq} \nabla_p \bar U_q  = {1 \over 2}
\bar A_1 (2 S_{ijk}^{(1)} + S_{ijk}^{(5)}) \;,
\nonumber\\
\varepsilon_{ijq} \bar K_{mm} \nabla_k \bar U_q  &=& {1 \over 2}
(3 \bar A_1 + \bar A_2) (2 S_{ijk}^{(1)} - S_{ijk}^{(5)}) \;,
\quad (\varepsilon_{ijq} \delta_{kp} + \varepsilon_{ijp}
\delta_{kq}) \bar K_{mm} \nabla_p \bar U_q = 2 (3 \bar A_1 + \bar
A_2) S_{ijk}^{(1)} \;,
\label{G1}
\end{eqnarray}
\end{widetext}
\noindent
where
\begin{eqnarray*}
S_{ijk}^{(1)} &=& \varepsilon_{ijp} (\partial \bar U)_{pk} \;,
\quad S_{ijk}^{(2)} = \varepsilon_{ikp} (\partial \bar U)_{pj} \;,
\\
S_{ijk}^{(3)} &=& \varepsilon_{jkp} (\partial \bar U)_{pi} \;,
\quad S_{ijk}^{(4)} = \bar W_{k} \delta_{ij} \;, \quad
S_{ijk}^{(5)} = \bar W_{j} \delta_{ik} \;,
\\
S_{ijk}^{(6)} &=& \varepsilon_{ikp} \beta_{jq} (\partial \bar
U)_{pq} \;, \quad S_{ijk}^{(7)} = \bar W_{k} \beta_{ij} \; .
\end{eqnarray*}
After the integration in Eq.~(\ref{S7}), we obtain
\begin{eqnarray}
{\cal E}^\sigma_{i} = b_{ijk}^{\sigma} \bar B_{j,k} \;,
\label{S8}
\end{eqnarray}
where the tensor $b_{ijk}^{\sigma}$ is given by
\begin{eqnarray}
b_{ijk}^{\sigma} = l_0^2 \, \biggl[\sum_{n=1}^7 D_n \,
S_{ijk}^{(n)} \biggr] \;,
\label{S9}
\end{eqnarray}
the coefficient $ D_3 = 0 $, and the other coefficients calculated
for $q=5/3$ are given by
\begin{widetext}
\begin{eqnarray}
D_1 &=& {1 \over 3} \biggl[A_1^{(2)} - 3 A_2^{(2)} - 18 C_1^{(2)}
+ \epsilon \biggl(A_1^{(2)} + A_2^{(2)} + {2 \over 3} C_1^{(2)}
\biggr) + \Psi_1 \biggl\{A_1 + 2 A_2 + {22 \over 3} C_1 - \epsilon
(2 A_1 + A_2 + 6 C_1) \biggr\}
\nonumber\\
&& + \Psi_2 \biggl\{- A_1 + {1 \over 3} C_1 + \epsilon \biggl(A_1
- {11 \over 3} C_1 \biggr) \biggr\} - (1 - \epsilon) \Psi_3\{C_1\}
- \Psi_0\{2 A_1 - 3 C_1\} \biggr] \;,
\nonumber\\
D_2 &=& {1 \over 3} \biggl[-(A_1^{(2)} + A_2^{(2)} + 4 C_1^{(2)})
+ \epsilon \biggl(- A_1^{(2)} + A_2^{(2)} + {32 \over 3} C_1^{(2)}
\biggr) + \Psi_1 \biggl\{- A_1 + A_2 + {70 \over 3} C_1 - 2
\epsilon (A_2 + 19 C_1) \biggr\}
\nonumber\\
&& + \Psi_2 \biggl\{A_1 - {71 \over 3} C_1 - \epsilon \biggl(A_1 -
{79 \over 3} C_1 \biggr) \biggr\} + (1 - \epsilon)
\biggl(\Psi_3\{- 2 A_1 + 7 C_1\} - {16 \over 3} \Psi_4\{C_1\} + 8
\Psi_5\{C_1\} \biggr)
\nonumber\\
&& + \Psi_0 \biggl\{2 A_1 - {11 \over 3} C_1 \biggr\} \biggr] \;,
\nonumber\\
D_4 &=& {1 \over 6} \biggl[3 A_1^{(2)} + A_2^{(2)} - {14 \over 3}
C_1^{(2)} + \epsilon \biggl(3 A_1^{(2)} - A_2^{(2)} - {26 \over 3}
C_1^{(2)} \biggr) - \Psi_1 \biggl\{A_1 + A_2 - {4 \over 3} C_1 - 2
\epsilon \biggl(A_1 + A_2
\nonumber\\
&& + {2 \over 3} C_1 \biggr) \biggr\} + (1 - \epsilon)
\biggl(\Psi_2\{A_1 + C_1\} - \Psi_3\{C_1\} \biggr) + \Psi_0
\{C_1\} \biggr] \;,
\nonumber\\
D_5 &=& {1 \over 6} \biggl[A_1^{(2)} + A_2^{(2)} - {14 \over 3}
C_1^{(2)} + \epsilon \biggl(A_1^{(2)} - A_2^{(2)} - {26 \over 3}
C_1^{(2)} \biggr) - \Psi_1 \biggl\{A_1 - A_2 - {4 \over 3} C_1 - 2
\epsilon \biggl(A_1 - A_2
\nonumber\\
&& + {2 \over 3} C_1 \biggr) \biggr\} + (1 - \epsilon)
\biggl(\Psi_2\{A_1 + C_1\} - \Psi_3\{C_1\} \biggr) + \Psi_0
\{C_1\} \biggr] \;,
\nonumber\\
D_6 &=& {1 \over 3} \biggl[A_2^{(2)} - 4 C_3^{(2)} - \epsilon
\biggl(A_2^{(2)} - {32 \over 3} C_3^{(2)} \biggr) + \Psi_1
\biggl\{- 3 A_2 + {70 \over 3} C_3 + 2 \epsilon (A_2 - 19 C_3)
\biggr\} - {1 \over 3} (71 - 79 \epsilon) \Psi_2 \{C_3\}
\nonumber\\
&& - (1 - \epsilon) \biggl(\Psi_3\{A_2 - 7 C_3\} + {16 \over 3}
\Psi_4\{C_3\} - 8 \Psi_5\{C_3\} \biggr) + \Psi_0 \biggl\{A_2 - {11
\over 3} C_3 \biggr\} \biggr] \;,
\nonumber\\
D_7 &=& {1 \over 6} \biggl[A_2^{(2)} - {14 \over 3} C_3^{(2)} +
\epsilon \biggl(3 A_2^{(2)} - {26 \over 3} C_3^{(2)} \biggr) +
\Psi_1 \biggl\{A_2 + {4 \over 3} (1 + \epsilon) C_3\biggr\} + (1 -
\epsilon) \biggl(\Psi_2 \{2 A_2 + C_3\}
\nonumber\\
&& - \Psi_3\{A_2 + C_3\}\biggr) + \Psi_0 \{A_2 + C_3\} \biggr]
\; . \label{G2}
\end{eqnarray}
\end{widetext}
\noindent The functions $\bar A_{k}(y)$ and $\bar C_{k}(y)$ are
determined by Eqs.~(\ref{P22}) in Appendix B, and the functions
$A_{k}^{(2)}(y)$ and $C_{k}^{(2)}(y)$ are determined by
Eqs.~(\ref{P21}) in Appendix D. The functions $\Psi_k\{X\}$ are
given by
\begin{eqnarray}
\Psi_0\{X\} &=& - {1 \over 2} (1 + \epsilon) X^{(2)}(0) + (2 -
\epsilon) X^{(2)}(\sqrt{2}\beta)
\nonumber\\
&& - {3 \over 4 \pi} (1 - \epsilon) \bar X(2 \beta^2) \;,
\nonumber \\
\Psi_1\{X\} &=& - 3 X^{(2)}(\sqrt{2}\beta) + {3 \over 2 \pi} \bar
X(2 \beta^2) \;,
\nonumber \\
\Psi_2\{X\} &=& 3 X^{(2)}(\sqrt{2}\beta) - {3 \over 4 \pi}
\biggl[2 \bar X(y) + y \bar X'(y) \biggr]_{y=2\beta^2} \;,
\nonumber\\
\Psi_3\{X\} &=& - 6 X^{(2)}(\sqrt{2}\beta) + {3 \over 4 \pi}
\biggl[4 \bar X(y) + y \bar X'(y) \biggr]_{y=2\beta^2} \;,
\nonumber\\
\Psi_4\{X\} &=& 4 X^{(2)}(\sqrt{2}\beta) - {1 \over 4 \pi}
\biggl[8 \bar X(y) + 4 y \bar X'(y)
\nonumber\\
&& + y^2 \bar X''(y) \biggr]_{y=2\beta^2} \;,
\nonumber\\
\Psi_5\{X\} &=& - {1 \over 2} X^{(2)}(\sqrt{2}\beta) + {1 \over 8
\pi} \biggl[2 \bar X(y) + y \bar X'(y)
\nonumber\\
&& + 2 y^2 \bar X''(y) \biggr]_{y=2\beta^2} \; .
\label{G3}
\end{eqnarray}
In Eqs.~(\ref{G1})-(\ref{G3}) we took into account that for the
''shear-current" dynamo, $\bar B_x /\bar B_y \sim l_0/L_B \ll 1 ,$
where $L_B$ is the characteristic scale of the mean magnetic field
variations. The nonlinear coefficient defining the
''shear-current'' effect is determined by
\begin{eqnarray}
\sigma_0(\bar{B}) &=& {1 \over 2} (D_2 + 2 D_4 + D_6 + 2 D_7) \; .
\label{S30}
\end{eqnarray}
Equation~(\ref{S30}) yields
\begin{eqnarray}
\sigma_0(\bar{B}) = \phi_1\{A_1 + A_2\} + \phi_2\{C_1 + C_3\} \;,
\label{F1}
\end{eqnarray}
where
\begin{eqnarray}
\phi_1\{X\} &=& {1 \over 3} \biggl[(1 + \epsilon)
X^{(2)}(\sqrt{2}\beta) + [\Psi_0 - (1 - \epsilon)(\Psi_1
\nonumber\\
&& - \Psi_2 + \Psi_3)]\{X\} \biggr] \;,
\label{G10} \\
\phi_2\{X\} &=& {1 \over 9} \biggl[(3 \epsilon - 13)
X^{(2)}(\sqrt{2}\beta) + [4 \Psi_2 - 4\Psi_0
\nonumber\\
&& - 18 \Psi_1 + (1 - \epsilon)(55 \Psi_1 - 38 \Psi_2 + 9 \Psi_3
\nonumber\\
&& - 8 \Psi_4 + 12 \Psi_5)]\{X\} \biggr] \; .
\label{G11}
\end{eqnarray}
The nonlinear dependence of the parameter $\sigma_0(\bar{B})$
determined by Eq.~(\ref{F1}), is shown in FIG.~3 for different
values of the parameter $\epsilon$. The asymptotic formula for the
parameter $\sigma_0(\bar{B})$ for $\bar{B} \ll \bar{B}_{\rm eq} /
4$ and $\bar{B} \gg \bar{B}_{\rm eq} / 4$ are given by
Eqs.~(\ref{F2})-(\ref{F4}). For the derivation of Eq.~(\ref{F1})
we used identities~(\ref{G20}) in Appendix D.

\section{The identities used for the integration in $ {\bf k} $--space}

To integrate over the angles in $ {\bf k} $--space we used the
following identities:
\begin{widetext}
\begin{eqnarray}
\bar K_{ij} &=& \int {k_{ij} \sin \theta \over 1 + a \cos^{2}
\theta} \,d\theta \,d\varphi = \bar A_{1} \delta_{ij} + \bar A_{2}
\beta_{ij} \;,
\label{C22} \\
\bar K_{ijmn} &=& \int {k_{ijmn} \sin \theta \over 1 + a \cos^{2}
\theta} \,d\theta \,d\varphi = \bar C_{1} (\delta_{ij} \delta_{mn}
+ \delta_{im} \delta_{jn} + \delta_{in} \delta_{jm}) + \bar C_{2}
\beta_{ijmn} + \bar C_{3} (\delta_{ij} \beta_{mn} + \delta_{im}
\beta_{jn}
\nonumber \\
&&+ \delta_{in} \beta_{jm} + \delta_{jm} \beta_{in} + \delta_{jn}
\beta_{im} + \delta_{mn} \beta_{ij}) \;,
\label{C24}\\
\bar H_{ijmn}(a) &=& \int {k_{ijmn} \sin \theta \over (1 + a
\cos^{2} \theta)^{2} } \,d\theta \,d\varphi = - \biggl( {\partial
\over \partial b } \int {k_{ijmn} \sin \theta \over b + a \cos^{2}
\theta} \,d\theta \,d\varphi \biggr)_{b=1} = \bar K_{ijmn}(a) + a
{\partial \over \partial a} \bar K_{ijmn}(a) \;,
\label{C23}\\
\bar G_{ijmn}(a) &=& \int {k_{ijmn} \sin \theta \over (1 + a
\cos^{2} \theta)^{3} } \,d\theta \,d\varphi = - {1 \over 2}
\biggl( {\partial \over \partial b } \int {k_{ijmn} \sin \theta
\over (b + a \cos^{2} \theta)^2} \,d\theta \,d\varphi
\biggr)_{b=1} = \bar H_{ijmn}(a) + {a \over 2} {\partial \over
\partial a} \bar H_{ijmn}(a) \;,
\label{CC25}\\
\bar Q_{ijmn}(a) &=& \int {k_{ijmn} \sin \theta \over (1 + a
\cos^{2} \theta)^{4} } \,d\theta \,d\varphi = - {1 \over 3}
\biggl( {\partial \over \partial b } \int {k_{ijmn} \sin \theta
\over (b + a \cos^{2} \theta)^3} \,d\theta \,d\varphi
\biggr)_{b=1} = \bar G_{ijmn}(a) + {a \over 3} {\partial \over
\partial a} \bar G_{ijmn}(a) \;,
\label{C25}
\end{eqnarray}
\end{widetext}
\noindent
where $ a = [\beta u_{0} k \tau(k) / 2]^{2} ,$ $ \hat
\beta_{i} = \beta_{i} / \beta ,$ $ \beta_{ij} = \hat \beta_{i}
\hat \beta_{j} ,$ and
\begin{eqnarray}
\bar A_{1} &=& {2 \pi \over a} \biggl[(a + 1) {\arctan (\sqrt{a}) \over
\sqrt{a}} - 1 \biggr] \;,
\nonumber\\
\bar A_{2} &=& - {2 \pi \over a} \biggl[(a + 3) {\arctan (\sqrt{a}) \over
\sqrt{a}} - 3 \biggr] \;,
\nonumber\\
\bar C_{1} &=& {\pi \over 2a^{2}} \biggl[(a + 1)^{2} {\arctan
(\sqrt{a}) \over \sqrt{a}} - {5 a \over 3} - 1 \biggr]  \;,
\nonumber\\
\bar C_{2} &=& \bar A_{2} - 7 \bar A_{1} + 35 \bar C_{1} \;,
\nonumber\\
\bar C_{3} &=& \bar A_{1} - 5 \bar C_{1} \; .
\label{P22}
\end{eqnarray}
In the case of $ a \ll 1 $ these functions are given by
\begin{eqnarray*}
\bar A_{1}(a) &\sim& {4 \pi \over 3} \biggl(1 - {1 \over 5} a
\biggr) \;, \quad \bar A_{2}(a) \sim - {8 \pi \over 15} a \;,
\\
\bar C_{1}(a) &\sim& {4 \pi \over 15} \biggl(1 - {1 \over 7} a
\biggr) \;, \quad \bar C_{2}(a) \sim \sim {32 \pi \over 315} a^{2}
\;,
\\
\bar C_{3}(a) &\sim& - {8 \pi \over 105} a \; .
\end{eqnarray*}
In the case of $ a \gg 1 $ these functions are given by
\begin{eqnarray*}
\bar A_{1}(a) &\sim& {\pi^{2} \over \sqrt{a}} - {4 \pi \over a}
\;, \quad \bar A_{2}(a) \sim - {\pi^{2} \over \sqrt{a}} + {8 \pi
\over a} \;,
\\
\bar C_{1}(a) &\sim& {\pi^{2} \over 4 \sqrt{a}} - {4 \pi \over 3
a} \;, \quad \bar C_{2}(a) \sim {3 \pi^{2} \over 4 \sqrt{a}} - {32
\pi \over 3 a} \;,
\\
\bar C_{3}(a) &\sim& - {\pi^{2} \over 4 \sqrt{a}} + {8 \pi \over 3
a} \; .
\end{eqnarray*}

\section{The functions $ A_{\alpha}^{(1)}(\beta) $ and
$ C_{\alpha}^{(1)}(\beta) $}

The functions $A_{n}^{(1)}(\beta)$ are defined as
\begin{eqnarray*}
A_{n}^{(1)}(\beta) = {3 \beta^{4} \over \pi} \int_{\beta}^{\infty}
{\bar A_{n}(X^{2}) \over X^{5}} \,d X  \;,
\end{eqnarray*}
and similarly for $ C_{n}^{(1)}(\beta) ,$ where
\begin{eqnarray*}
X^{2} = \beta^{2} (k / k_{0})^{2/3} = a = [\beta u_{0} k \tau(k) /
2]^{2}  \;,
\end{eqnarray*}
and we took into account that the inertial range of the turbulence
exists in the scales: $ l_{d} \leq r \leq l_{0} .$ Here the
maximum scale of the turbulence $ l_{0} \ll L_{B} ,$ and $ l_{d} =
l_{0} / {\rm Re}^{3/4} $ is the viscous scale of turbulence, and $
L_{B} $ is the characteristic scale of variations of the
nonuniform mean magnetic field. For very large Reynolds numbers $
k_{d} = l_{d}^{-1} $ is very large and the turbulent hydrodynamic
and magnetic energies are very small in the viscous dissipative
range of the turbulence $ 0 \leq r \leq l_{d} .$ Thus we
integrated in $ \bar A_{n} $ over $ k $ from $ k_{0} = l_{0}^{-1}
$ to $ \infty .$ We also used the following identity
\begin{eqnarray*}
\int_0^1 \bar A_{n}(a(\bar \tau)) \bar \tau \, d\bar \tau = {2 \pi
\over 3} A_{n}^{(1)}(\beta)   \;,
\end{eqnarray*}
and similarly for $ C_{n}^{(1)}(\beta)$. The functions $
A_{\alpha}^{(1)}(\beta) $ and $ C_{\alpha}^{(1)}(\beta) $ are
given by
\begin{eqnarray}
A_{1}^{(1)}(\beta) &=& {6 \over 5} \biggl[{\arctan \beta \over
\beta} \biggl(1 + {5 \over 7 \beta^{2}} \biggr) + {1 \over 14}
L(\beta) - {5 \over 7\beta^{2}} \biggr]  \;,
\nonumber\\
A_{2}^{(1)}(\beta) &=& - {6 \over 5} \biggl[{\arctan \beta \over
\beta} \biggl(1 + {15 \over 7 \beta^{2}} \biggr) - {2 \over 7}
L(\beta) - {15 \over 7\beta^{2}} \biggr]  \;,
\nonumber\\
C_{1}^{(1)}(\beta) &=& {3 \over 10} \biggl[{\arctan \beta \over
\beta} \biggl(1 + {10 \over 7 \beta^{2}} + {5 \over 9 \beta^{4}}
\biggr) + {2 \over 63} L(\beta)
\nonumber \\
& & - {235 \over 189 \beta^{2}} - {5 \over 9 \beta^{4}} \biggr]
\;,
\nonumber\\
C_{2}^{(1)}(\beta) &=& A_{2}^{(1)}(\beta) - 7 A_{1}^{(1)}(\beta) +
35 C_{1}^{(1)}(\beta) \;,
\nonumber\\
C_{3}^{(1)}(\beta) &=& A_{1}^{(1)}(\beta) - 5 C_{1}^{(1)}(\beta)
\label{P20} \;,
\end{eqnarray}
where $ L(\beta) = 1 - 2 \beta^{2} + 2 \beta^{4} \ln (1 + \beta^{-2}) .$
For $ \beta \ll 1 $ these functions are given by
\begin{eqnarray*}
A_{1}^{(1)}(\beta) &\sim& 1 - {2 \over 5} \beta^{2}  \;, \quad
A_{2}^{(1)}(\beta) \sim - {4 \over 5} \beta^{2} \;,
\\
C_{1}^{(1)}(\beta) &\sim& {1 \over 5}  \biggl(1 - {2 \over 7}
\beta^{2}\biggr) \;, \quad C_{2}^{(1)}(\beta) \sim - {32 \over
105} \beta^{4} \ln \beta \;,
\\
C_{3}^{(1)}(\beta) &\sim& - {4 \over 35} \beta^{2} \;,
\end{eqnarray*}
and for $ \beta \gg 1 $ they are given by
\begin{eqnarray*}
A_{1}^{(1)}(\beta) &\sim& {3 \pi \over 5 \beta} - {2 \over
\beta^{2}} \;, \quad A_{2}^{(1)}(\beta) \sim - {3 \pi \over 5
\beta} + {4 \over \beta^{2}} \;,
\\
C_{1}^{(1)}(\beta) &\sim& {3 \pi \over 20 \beta} - {2 \over
3\beta^{2}} \;, \quad C_{2}^{(1)}(\beta) \sim {9 \pi \over 20
\beta} \;,
\\
C_{3}^{(1)}(\beta) &\sim& - {3 \pi \over 20 \beta} + {4 \over 3
\beta^{2}} \; .
\end{eqnarray*}
Here we used that for $ \beta \ll 1 $ the function $ L(\beta) \sim
1 - 2 \beta^{2} - 4 \beta^{4} \ln \beta ,$ and for $ \beta \gg 1 $
the function $ L(\beta) \sim 2 / 3 \beta^{2} .$ We also use the
identity:
\begin{eqnarray*}
\int_0^1 \bar H_{ijmn}(a(\bar \tau)) \, \bar \tau^4 \, \biggl({k
\over k_0} \biggr)^2 \, d\bar \tau &=& 2 \pi K_{ijmn}^{(1)}(\beta)
\\
&& - \bar K_{ijmn}(\beta^2) \; .
\end{eqnarray*}

\section{The functions $ A_{\alpha}^{(2)}(\beta) $ and
$ C_{\alpha}^{(2)}(\beta) $}

The functions $A_{n}^{(2)}(\beta)$ are defined as
\begin{eqnarray*}
A_{n}^{(2)}(\beta) = {3 \beta^{6} \over \pi} \int_{\beta}^{\infty}
{\bar A_{n}(X^{2}) \over X^{7}} \,d X \;,
\end{eqnarray*}
and similarly for $ C_{n}^{(2)}(\beta) .$ We used the following
identity
\begin{eqnarray*}
\int_0^1 \bar A_{n}(a(\bar \tau)) \bar \tau^2 \, d\bar \tau = {2
\pi \over 3} A_{n}^{(2)}(\beta)  \;,
\end{eqnarray*}
and similarly for $ C_{n}^{(2)}(\beta) .$ The functions $
A_{\alpha}^{(2)}(\beta) $ and $ C_{\alpha}^{(2)}(\beta) $ are
given by
\begin{eqnarray}
A_{1}^{(2)}(\beta) &=& F(1; -1; 0) \;,
\nonumber\\
A_{2}^{(2)}(\beta) &=& F(-1; 3; 0) \;, \,
\nonumber\\
C_{1}^{(2)}(\beta)  &=& (1/4) F(1; -2; 1) \;, \,
\nonumber\\
C_{2}^{(2)}(\beta) &=& (1/4) F(3; -30; 35) \;, \,
\nonumber\\
C_{3}^{(2)}(\beta)  &=& (1/4) F(-1; 6; -5) \;,
\label{P21}
\end{eqnarray}
where
\begin{eqnarray*}
F(\alpha; \sigma; \gamma) = \pi [\alpha J_{0}^{(2)}(\beta) +
\sigma J_{2}^{(2)}(\beta) + \gamma J_{4}^{(2)}(\beta)] \;,
\end{eqnarray*}
\begin{eqnarray*}
J_{0}^{(2)}(\beta) &=& {1 \over 7 \pi} \biggl(1 + 6 {\arctan \beta
\over \beta} - {3 \beta^{2} \over 2} L(\beta) \biggr) \;,
\\
J_{2}^{(2)}(\beta) &=& {7 \over 9} J_{0}^{(2)}(\beta) + \tilde
L(\beta) \;,
\\
J_{4}^{(2)}(\beta) &=& {9 \over 11} \biggl(J_{2}^{(2)}(\beta) - {1
\over \beta^{2}} \tilde L(\beta) - {4 \over 9 \pi \beta^{2}}
\biggr)\;,
\\
\tilde L(\beta) &=& {2 \over 3 \pi \beta^{2}}  \biggl(1 - {\arctan
\beta \over \beta} (1 + \beta^{2}) \biggr) \; .
\end{eqnarray*}
For $ \beta \ll 1 $ the functions $ J_{\alpha}^{(2)}(\beta) $ are
given by
\begin{eqnarray*}
J_{0}^{(2)}(\beta) &\sim& {1 \over \pi} \biggl(1 - {1 \over 2}
\beta^{2}\biggr) \;,
\\
J_{2}^{(2)}(\beta) &\sim& {1 \over 3\pi} \biggl(1 - {9 \over 10}
\beta^{2} \biggr) \;,
\\
J_{4}^{(2)}(\beta) &\sim& {1 \over 5\pi} \biggl(1 - {15 \over 14}
\beta^{2} \biggr) \;,
\end{eqnarray*}
and for $ \beta \gg 1 $ they are given by
\begin{eqnarray*}
J_{0}^{(2)}(\beta) &\sim& {3 \over 7 \beta} - {3 \over 4 \pi
\beta^2} \;,
\\
J_{2}^{(2)}(\beta) &\sim& {3 \over 4 \pi \beta^2} \;,
\\
J_{4}^{(2)}(\beta) &\sim& {1 \over 4 \pi \beta^2} \; .
\end{eqnarray*}
For $ \beta \ll 1 $ the functions $ A_{\alpha}^{(2)}(\beta) $ and
$ C_{\alpha}^{(2)}(\beta) $ are given by
\begin{eqnarray*}
A_{1}^{(2)}(\beta) &\sim& {2 \over 3}\biggl(1 - {3 \over 10}
\beta^{2}\biggr) \;, \quad A_{2}^{(2)}(\beta) \sim - {2 \over 5}
\beta^{2} \;,
\\
C_{1}^{(2)}(\beta) &\sim& {2 \over 15}  \biggl(1 - {3 \over 14}
\beta^{2}\biggr) \;, \quad C_{2}^{(2)}(\beta) \sim O(\beta^{4})
\;,
\\
C_{3}^{(2)}(\beta) &\sim& - {2 \over 35} \beta^{2} \;,
\end{eqnarray*}
and for $ \beta \gg 1 $ they are given by
\begin{eqnarray*}
A_{1}^{(2)}(\beta) &\sim& {3 \pi \over 7 \beta} - {3 \over 2
\beta^2} \;, \quad A_{2}^{(2)}(\beta) \sim - {3 \pi \over 7 \beta}
+ {3 \over \beta^2} \;,
\\
C_{1}^{(2)}(\beta) &\sim& {3 \pi \over 28 \beta} - {1 \over 2
\beta^2} \;, \quad C_{2}^{(2)}(\beta) \sim {9 \pi \over 28 \beta}
- {4 \over \beta^2}\;,
\\
C_{3}^{(2)}(\beta) &\sim& - {3 \pi \over 28 \beta} + {1 \over
\beta^2} \; .
\end{eqnarray*}

We also used the following identities:
\begin{eqnarray}
\Psi_1\{K_{ijmn}\} &=& K_{ijmn}^{(2)}(\sqrt{2}\beta) -
H_{ijmn}^{(2)}(\sqrt{2}\beta) \;,
\nonumber \\
\Psi_2\{K_{ijmn}\} &=& K_{ijmn}^{(2)}(\sqrt{2}\beta) - 2
H_{ijmn}^{(2)}(\sqrt{2}\beta)
\nonumber\\
&& + G_{ijmn}^{(2)}(\sqrt{2}\beta) \;,
\nonumber\\
\Psi_3\{K_{ijmn}\} &=& H_{ijmn}^{(2)}(\sqrt{2}\beta) -
G_{ijmn}^{(2)}(\sqrt{2}\beta) \;,
\nonumber\\
\Psi_4\{K_{ijmn}\} &=& H_{ijmn}^{(2)}(\sqrt{2}\beta) - 2
G_{ijmn}^{(2)}(\sqrt{2}\beta)
\nonumber\\
&& + Q_{ijmn}^{(2)}(\sqrt{2}\beta) \;,
\nonumber\\
\Psi_5\{K_{ijmn}\} &=& {1 \over 2}
\biggl[K_{ijmn}^{(2)}(\sqrt{2}\beta) - 3
H_{ijmn}^{(2)}(\sqrt{2}\beta)
\nonumber\\
&& + 3 G_{ijmn}^{(2)}(\sqrt{2}\beta) -
Q_{ijmn}^{(2)}(\sqrt{2}\beta) \biggr] \;,
\nonumber\\
\label{G20}
\end{eqnarray}
where
\begin{eqnarray*}
H_{ijmn}^{(2)}(\sqrt{2}\beta) &=& 4 K_{ijmn}^{(2)}(\sqrt{2}\beta)
- {3 \over 2 \pi} \bar K_{ijmn}(2\beta^2) \;,
\\
G_{ijmn}^{(2)}(\sqrt{2}\beta) &=& {5 \over 2}
H_{ijmn}^{(2)}(\sqrt{2}\beta) - {3 \over 4 \pi} \bar
H_{ijmn}(2\beta^2)\;,
\\
Q_{ijmn}^{(2)}(\sqrt{2}\beta) &=& 2 G_{ijmn}^{(2)}(\sqrt{2}\beta)
- {1 \over 2 \pi} \bar G_{ijmn}(2\beta^2) \; .
\end{eqnarray*}


\begin{thebibliography}{}

\bibitem {M78} H. K. Moffatt, {\em Magnetic Field Generation in
Electrically Conducting  Fluids} (Cambridge University Press, New
York, 1978).

\bibitem {P79} E. Parker, {\it Cosmical Magnetic Fields} (Oxford
University Press,  New York,  1979).

\bibitem {KR80} F. Krause, and K. H. R\"{a}dler, {\it Mean-Field
Magnetohydrodynamics and  Dynamo Theory} (Pergamon, Oxford, 1980).

\bibitem {ZRS83} Ya. B. Zeldovich, A. A. Ruzmaikin, and D. D.
Sokoloff, {\em Magnetic Fields in Astrophysics} (Gordon and
Breach, New York, 1983).

\bibitem {RSS88} A. Ruzmaikin, A. M. Shukurov, and D. D.
Sokoloff, {\it Magnetic Fields of Galaxies} (Kluwer Academic,
Dordrecht, 1988).

\bibitem {S89} M. Stix, {\it The Sun: An Introduction} (Springer,
Berlin and Heidelberg, 1989).

\bibitem {RS92}
P. H. Roberts and A. M. Soward, Annu. Rev. Fluid Mech. {\bf 24},
459 (1992), and references therein.

\bibitem{BB96}
R. Beck,  A. Brandenburg, D. Moss , A. Shukurov and D. Sokoloff,
Ann. Rev. Astron. Astrophys. {\bf 34}, 155 (1996), and references
therein.

\bibitem{K99}
R. Kulsrud, Ann. Rev. Astron. Astrophys. {\bf 37}, 37 (1999), and
references therein.

\bibitem {B2000}
F. H. Busse, Annu. Rev. Fluid Mech. {\bf 32}, 383 (2000), and
references therein.

\bibitem {R69} K.-H. R\"{a}dler,
Monatsber. Dtsch. Akad. Wiss. Berlin {\bf 11}, 272 (1969).

\bibitem {R72} P.H. Roberts, Phil. Trans. R. Soc. London
{\bf A 272}, 663 (1972).

\bibitem {MP82} H.K. Moffatt and M.R.E. Proctor,
Geophys. Astrophys. Fluid Dyn. {\bf 21}, 265 (1982).

\bibitem {R86} K.-H. R\"{a}dler,  Astron. Nachr. {\bf 307}, 89
(1986).

\bibitem {RKR03}
K.-H. R\"{a}dler, N. Kleeorin and I. Rogachevskii, Geophys.
Astrophys. Fluid Dyn. {\bf 97}, 249 (2003).

\bibitem {RK03}
I. Rogachevskii and N. Kleeorin, Phys. Rev. E {\bf 68}, 036301
(2003).

\bibitem {R80} K.-H. R\"{a}dler,  Astron. Nachr. {\bf 301}, 101
(1980); Geophys. Astrophys. Fluid Dynamics {\bf 20}, 191 (1982).

\bibitem {RS02} K.-H. R\"{a}dler and R. Stepanov, Proc. 5th
International PAMIR Conference on Fundamental and Applied MHD,
16-20 September 2002, Ramatuelle, France, v. 2, VI 77 - VI 82
(2002).

\bibitem {KR82} N. Kleeorin, and A. Ruzmaikin,
Magnetohydrodynamics {\bf No. 2}, 17 (1982).

\bibitem {KRR94} N. Kleeorin, I. Rogachevskii, and A. Ruzmaikin,
Solar Phys. {\bf 155}, 223 (1994); Astron. Astrophys. {\bf 297},
159 (1995).

\bibitem {GD94} A. Gruzinov, and P. H. Diamond, Phys.
Rev. Lett., {\bf 72}, 1651 (1994); Phys. Plasmas {\bf 2}, 1941
(1995).

\bibitem {KR99}
N. Kleeorin and I. Rogachevskii, Phys. Rev. E {\bf 59}, 6724
(1999).

\bibitem {KMRS2000}
N. Kleeorin, D. Moss, I. Rogachevskii and D. Sokoloff, Astron.
Astrophys. {\bf 361}, L5 (2000); {\bf 387}, 453 (2002); {\bf 400},
9 (2003).

\bibitem {BB02} E.G. Blackman and A. Brandenburg, Astrophys. J.
{\bf 579}, 359 (2002).

\bibitem {MY75} A. S. Monin and A. M. Yaglom, {\it Statistical Fluid
Mechanics}  (MIT Press, Cambridge, Massachusetts, 1975), Vol. 2.

\bibitem {Mc90} W. D. McComb, {\it The Physics of Fluid Turbulence}
(Clarendon,  Oxford, 1990).

\bibitem {O70}  S. A. Orszag, J. Fluid Mech. {\bf 41}, 363 (1970).

\bibitem {PFL76} A. Pouquet, U. Frisch, and J. Leorat, J. Fluid Mech.
{\bf 77}, 321 (1976).

\bibitem {KRR90}
N. Kleeorin, I. Rogachevskii, and A. Ruzmaikin, Zh. Eksp. Teor.
Fiz. {\bf 97}, 1555 (1990) [Sov. Phys. JETP {\bf 70}, 878 (1990)].

\bibitem {KMR96} N. Kleeorin, M. Mond, and I. Rogachevskii,
Astron. Astrophys. {\bf 307}, 293 (1996).

\bibitem {RK2000}
I. Rogachevskii and N. Kleeorin, Phys. Rev. E {\bf 61}, 5202
(2000); ibid {\bf 64}, 056307 (2001).

\bibitem {KR03}
N. Kleeorin and I. Rogachevskii, Phys. Rev. E {\bf 67}, 026321
(2003).

\bibitem {BK04}
A. Brandenburg, P. K\"{a}pyl\"{a}, and A. Mohammed, Phys. Fluids
{\bf 16}, 1020 (2004).

\bibitem {FB99}
G. B. Field, E. G. Blackman and H. Chou, Astrophys. J. {\bf 513},
638 (1999).

\bibitem {RS75} P. H. Roberts and A. M. Soward,  Astron. Nachr. {\bf
296}, 49 (1975).

\bibitem {KR94}
N. Kleeorin and I. Rogachevskii, Phys. Rev. E {\bf 50}, 2716
(1994).

\end{thebibliography}
\end{document}